\documentclass[aps,prd,reprint,nofootinbib]{revtex4}

\usepackage{graphicx}
\usepackage{dcolumn}
\usepackage{bm}
\usepackage{amsmath}
\usepackage{wrapfig}
\usepackage{amsmath,amssymb,mathrsfs}
\usepackage{latexsym}
\usepackage{graphicx}
\usepackage{subfigure}
\usepackage{slashed}
\usepackage{epstopdf}
\usepackage{color}
\usepackage{ulem}
\usepackage{float}
\usepackage{multirow}
\usepackage{bm}
\usepackage{simplewick}
\newcommand{\RI}{\textrm{RI}}
\newcommand{\msbar}{\overline{\textrm{MS}}}
\newcommand{\mmsbar}{\textrm{M}\overline{\textrm{MS}}}
\newcommand{\MSb}{\overline{\textrm{MS}}}
\newcommand{\MMSb}{\textrm{M}\overline{\textrm{MS}}}

\def\PKU{a}
\def\cyp{b}
\def\cyi{c}
\def\poz{d}
\def\teml{e}
\def\CICQ{f}
\def\CHEP{g}
\def\LNPT{h}
\def\desy{i}
\def\bon{j}
\begin{document}


%
%
%

  \begin{center}
    \begin{LARGE}
            \textbf{Parton distribution functions of $\Delta^+$ on the lattice}
    \end{LARGE}
  \end{center}

  \baselineskip 20pt plus 2pt minus 2pt
  \begin{center}
    \textbf{
      Yahui Chai$^{(\PKU)}$,
      Yuan Li$^{(\PKU)}$,
      Shicheng Xia$^{(\PKU)}$,
      Constantia Alexandrou$^{(\cyp, \cyi)}$,
      Krzysztof Cichy$^{(\poz)}$,\\
      Martha Constantinou$^{(\teml)}$,
      Xu Feng$^{(\PKU,\CICQ,\CHEP,\LNPT)}$
      Kyriakos Hadjiyiannakou$^{(\cyi)}$,
      Karl Jansen$^{(\desy)}$,
      Giannis Koutsou$^{(\cyi)}$,
      Chuan Liu$^{(\PKU,\CICQ,\CHEP)}$
      Aurora Scapellato$^{(\poz)}$,
      Fernanda Steffens$^{(\bon)}$
      }
\end{center}

  \begin{center}
    \begin{footnotesize}
      \noindent
    $^{(\PKU)}$ School of Physics, Peking University, Beijing 100871, China\\	
 	$^{(\cyp)}$ Department of Physics, University of Cyprus, P.O. Box 20537, 1678 Nicosia, Cyprus\\	
 	$^{(\cyi)}$ Computation-based Science and Technology Research Center, The Cyprus Institute, 20 Kavafi Str., Nicosia 2121, Cyprus \\
 	$^{(\poz)}$  Faculty of Physics, Adam Mickiewicz University, ul.\ Uniwersytetu Pozna\'{n}skiego 2, 61-614 Pozna\'{n}, Poland\\
 	$^{(\teml)}$ Temple University,1925 N. 12th Street, Philadelphia, PA 19122-1801, USA \\
 	$^{(\CICQ)}$ Collaborative Innovation Center of Quantum Matter, Beijing 100871, China\\
    $^{(\CHEP)}$ Center for High Energy Physics, Peking University, Beijing 100871, China \\
    $^{(\LNPT)}$ State Key Laboratory of Nuclear Physics and Technology, Peking University, Beijing 100871, China\\
    $^{(\desy)}$ NIC, DESY, Platanenallee 6, D-15738 Zeuthen, Germany \\
 	$^{(\bon)}$ Institut f\"{u}r Strahlen- und Kernphysik, Rheinische Friedrich-Wilhelms-Universit\"{a}t Bonn,  Nussallee 14-16, 53115 Bonn\\
     \vspace{0.2cm}
    \end{footnotesize}
  \end{center}

\date{\today}

\begin{abstract}
	We perform a first calculation for the unpolarized parton distribution function of the $\Delta^+$ baryon using lattice QCD simulations within the framework of Large Momentum Effective Theory. Two ensembles of $N_f=2+1+1$ twisted mass fermions  are utilized with a pion mass of 270 MeV and 360 MeV, respectively. The baryon, which is treated as a stable single-particle state, is boosted with momentum $P_3$ with values $\{0.42,0.83,1.25\}$ GeV, and we utilize momentum smearing to improve the signal. The unpolarized parton distribution function of $\Delta^+$ is obtained using a non-perturbative renormalization and a one-loop formula for the matching, with encouraging precision. In particular, we compute the $\overline{d}(x)-\overline{u}(x)$ asymmetry and compare it with the same quantity in the nucleon, in a first attempt towards resolving the physical mechanism responsible for generating such asymmetry.
	\end{abstract}

\maketitle


\section{introduction}
Quantum Chromodynamics (QCD) is established as the fundamental theory which describes the strong interaction among
quarks and gluons, the basic constituents of all hadronic matter. As such, it is valid in a wide range of energy scales, from the hadronic regime, where non-perturbative effects dominate, to high energy, where perturbation theory is applicable. Thus, an appropriate formulation is necessary to address the highly non-perturbative dynamics of QCD at low energies. Lattice QCD is an ideal non-perturbative tool, which relies on a space-time discretization in Euclidean space, and is successfully used to describe the properties of fundamental particles. 
 
A prime example of very important non-perturbative physical quantities are the 
parton distribution functions (PDFs). They encode important information on 
the internal structure of hadrons, such as the distribution of the spin and momentum of the parent hadron among 
its constituents.
In principle, lattice QCD provides a systematically improvable theoretical framework for extracting the PDFs from first principles. However, since PDFs are defined on the light cone, a direct computation of these objects on a Euclidean lattice remained elusive for many years.
Still, there has been a large and successful activity of the lattice QCD community to compute Mellin moments of PDFs. If a large number of the Mellin moments were available, they could, in principle, be used to reconstruct the PDFs. 
However, this seems to be unreachable from lattice calculations, since the statistical noise of higher moments is increasing rapidly and power-divergent operator mixings would have to be taken into account for the fourth and higher moments.
Thus, lattice computations of the Mellin moments have been restricted mostly to the first or second moment only. 
For recent works of the Extended Twisted Mass Collaboration for computing nucleon form factors, moments of the proton and pion and nucleon charges, see e.g.\ Refs.~\cite{Abdel-Rehim:2015owa,Alexandrou:2019olr,Alexandrou:2019brg,Alexandrou:2019ali,Alexandrou:2018sjm,Oehm:2018jvm}.

More recent efforts focus on alternative methods to access PDFs, starting from the pioneering proposal of X.\ Ji, which is based on the Large Momentum Effective Theory (LaMET) \cite{Ji:2013dva,Ji:2014gla}. 
In this approach, instead of calculating the light-cone quark
correlations in a hadron, one computes matrix elements with quarks separated by a space-like distance within
a boosted hadron, with such computation being feasible within the scope of lattice QCD. Applying a Fourier transform on these spatial matrix elements yields, after a suitable non-perturbative renormalization, the so-called quasi-PDFs, which share the same infrared physics as the light-cone PDFs \cite{Xiong:2013bka,Ma:2014jla,Briceno:2017cpo,Ma:2017pxb}. 
Because of the equivalence in the infrared region, the quasi-PDFs can be matched to light-cone PDFs using perturbation theory, and their difference is suppressed for large enough momentum boost of the hadron. Currently, the matching is known to one-loop level and is valid up to power-suppressed $O(M^2/P^2_3, \Lambda^2_{\rm QCD}/P^2_3)$ corrections, where $M$ and $P_3$ are the
mass and momentum of the boosted hadron. The hadron mass corrections are known to all orders in $M^2/P_3^2$ \cite{Chen:2016utp}; the higher-twist corrections have not been calculated yet. 
Besides quasi-PDFs, other theoretical proposals to calculate PDFs in lattice QCD have been proposed and explored~\cite{Liu:1993cv,Aglietti:1998ur,Detmold:2005gg,Braun:2007wv,Ma:2014jla,Radyushkin:2016hsy,Ma:2017pxb,Radyushkin:2017cyf,Orginos:2017kos,Radyushkin:2017lvu,Chambers:2017dov,Karpie:2018zaz,Radyushkin:2018cvn,Bali:2018spj,Detmold:2018kwu,Sufian:2019bol,Liang:2019frk,Joo:2019jct,Joo:2019bzr,Bhat:2020ktg}, which are complementary to the LaMET approach.

Since Ji's proposal, considerable progress has been made. 
Several numerical investigations have been performed, see e.g.\ Refs.~\cite{Lin:2014zya,Alexandrou:2015rja,Chen:2016utp,Alexandrou:2016jqi,Zhang:2017bzy,Lin:2017ani,Chen:2017gck,Chen:2018fwa,Alexandrou:2018pbm,Lin:2018qky,Liu:2018uuj,Fan:2018dxu,Alexandrou:2018eet,Izubuchi:2019lyk,Bhattacharya:2020cen}.
Theoretical progress encompassed, in particular, proving the renormalizability of quasi-PDFs \cite{Ishikawa:2017faj,Ji:2017oey,Zhang:2018diq,Li:2018tpe}, developing their non-perturbative renormalization programme \cite{Alexandrou:2017huk,Green:2017xeu} and calculating the one-loop matching coefficient~\cite{Xiong:2013bka,Ma:2014jla,Ji:2015qla,Xiong:2015nua,Ma:2017pxb,Wang:2017qyg,Stewart:2017tvs,Izubuchi:2018srq,Alexandrou:2018pbm,Alexandrou:2018eet,Liu:2018uuj,Liu:2018hxv}. For an extensive summary of different activities related to the quasi-distribution approach, as well as other approaches, see the review in Ref.~\cite{Cichy:2018mum}.

Although these developments show the success and promising potential of the LaMET approach to deliver
PDFs computed from first principles, the analyses of systematic effects, e.g.\ in Ref.~\cite{Alexandrou:2019lfo}, demonstrate that substantial amount of work still has to be performed to reach controlled results on a quantitative level. 
In particular, the presently reachable accuracy prevents addressing the important physical problem of understanding   
the sign and size of the antiquark asymmetry for light quarks in the nucleon sea, $\overline{d}(x) - \overline{u}(x)$, from a first-principle lattice QCD calculation of PDFs. 
Such a computation would constitute a major step forward to our understanding of the quark structure of the proton. Since this may become very difficult to achieve, given the limits that current lattice QCD calculation of PDFs have, an alternative way to test the origin
of the asymmetry is to study the quark structure of the $\Delta^+$ baryon, as proposed in Ref.~\cite{Ethier:2018efr}.

As in the case of the proton, the $\Delta^+$ baryon 
can oscillate into $N \pi$ and $\Delta \pi$. However, in contrast to
the proton and for kinematical reasons, there is a strong enhancement of
the light quark asymmetry in the $\Delta^+$, even if the $\Delta^+$ is treated as a stable state \cite{Ethier:2018efr}.
In particular, the predicted asymmetry
in the $\Delta^+$ is significantly larger, by about a factor of 2, than the asymmetry in the 
proton when the pion mass is only slightly larger than the mass difference between the Delta and the nucleon. 
The main motivation of our work originates from the role of
spontaneous chiral symmetry breaking in QCD to produce 
this asymmetry as an excess of $\bar{d}$ over $\bar{u}$, as discussed in Refs.~\cite{Thomas:2000ny,Chen:2001eg}. While it would be very hard to test this idea in fully physical conditions, where the $\Delta$ would be a resonance, in Ref.~\cite{Ethier:2018efr} an unphysical setup with the $\Delta$ being stable was provided which still allows to investigate the scenario that chiral symmetry breaking is responsible for the light quark asymmetry in the $\Delta$ sea. In particular, the setup of Ref.~\cite{Ethier:2018efr} can be used as a laboratory employing heavier than physical pion masses and in this paper, we will make use of this idea by working at pion masses of 270MeV and 360MeV as well as in a kinematical regime where the $\Delta$ can be considered a stable baryon.

It is the goal of this work to provide a very first lattice QCD study of the PDFs of the $\Delta$ baryon, with two ensembles leading to $m_\Delta=1.59(4),\,1.42(5)$ GeV. Since the  $\Delta$ baryon is heavier than the proton, it can be expected that the signal-to-noise problem is significantly enhanced. 
We show, however, that it is still possible to obtain PDFs in the $\Delta$ baryon with a good precision. 
The structure and quark content of the $\Delta(1232)$ baryon is presently not accessible experimentally. Lattice QCD is the only available framework to provide information on the resonance properties of the $\Delta$ baryon, using different techniques than for the nucleon, such as a full finite-volume multi-particle formalism~\cite{Briceno:2015tza,Baroni:2018iau}. Therefore, one can access information not only on the quark content of the $\Delta$, but also to evaluate its electromagnetic and axial form factors, as well as the $\Delta$-nucleon transition form factors. It is also an ideal formulation to study the unphysical case of a stable $\Delta$ baryon, as done in other theoretical studies, e.g., in Ref.~\cite{Ethier:2018efr}, which is the main scope in this work.

\section{Theoretical setup}
The unpolarized PDF, denoted by $q(x)$, is defined on the light cone as~\cite{Collins:2011zzd}
\begin{equation}
q(x)=\int_{-\infty}^{+\infty} \frac{d \xi^{-}}{4 \pi} e^{-i x P^{+} \xi^{-}}\left\langle h\left|\overline{\psi}\left(0\right) \gamma^{+} W\left(0,\xi^{-}\right) \psi(\xi^{-})\right| h\right\rangle ,
\label{def_unpolarized}
\end{equation}
where the light-cone vectors are taken as $\xi^{ \pm}=\left(\xi^{0} \pm \xi^{3}\right) / \sqrt{2}$. $W\left(0,\xi^{-}\right)=e^{-i g \int_{0}^{\xi^{-}} d \eta^{-} A^{+}\left(\eta^{-}\right)}$ is the Wilson line ensuring gauge invariance and $x$ is the momentum fraction
carried by the quarks in the hadron. The plus component of the  momentum $P^{+}$ is $\left(P^{0}+P^{3}\right) / \sqrt{2}$ and  $|h\rangle$ the hadron state of interest. Deep inelastic scattering physics is light-cone
dominated, meaning that $\xi^{2}=t^{2}-\vec{r}^{2}\approx 0$ in Minkowskian space. However, this condition is satisfied by only one single point (the origin) in Euclidean spacetime, making a direct extraction of PDFs impossible from lattice QCD using the definition in Eq.~(\ref{def_unpolarized}). Inspired by an equivalence between light-cone frame where the hadron is at rest and the infinite momentum frame (IMF), where the hadron is moving with  infinite momentum, the quasi-distribution approach proposes to extract PDFs from purely spatial correlation functions of as highly as possible boosted hadrons. Quasi-PDFs are then  defined by
\begin{equation}
\tilde{q}\left(x, P_3, \mu\right)=\int_{-\infty}^{+\infty} \frac{d z}{4 \pi} e^{-i x P_3 z}\langle h(P_3)|\overline{\psi}(0) \Gamma {W}(0,{z}) \psi(z)| h(P_3)\rangle,
\end{equation}
where $|h(P_3)\rangle$ is the boosted hadron state with four-momentum $P=\left(E, 0,0, P_3\right)$, and $W(0,z)$ is the Wilson line along the boosted direction, usually taken to be the $z$ direction on the lattice.
The scale $\mu$ is the energy scale at which renormalization is done. 
The Dirac structure, indicated by $\Gamma$, defines the type of PDF of interest. 
For the unpolarized PDF, $\Gamma$ is chosen to be $\gamma^0$ 
to avoid operator mixing for Wilson-type fermions \cite{Constantinou:2017sej}.
\par
Since the infrared physics is the same for quasi-PDFs and light-cone PDFs and the difference is only in 
the ultraviolet region \cite{Ma:2014jla,Ma:2017pxb}, the difference between the distributions can be computed in perturbation theory and quasi-PDFs are related to light-cone PDFs by 
the following matching equation
\begin{equation}
\tilde{q}\left(x, P_3, \mu\right)=\int_{-1}^{1} \frac{d y}{|y|} C\left(\frac{x}{y}, \frac{\mu}{yP_3}\right) q(y, \mu)+O\left(\frac{M^{2}}{P_3^{2}}, \frac{\Lambda_{Q C D}^{2}}{P_3^{2}}\right),
\label{eq:matching equation}
\end{equation}
within LaMET~\cite{Ji:2014gla}. In Eq.~(\ref{eq:matching equation}), $q(y,\mu)$ is the light-cone PDF at the scale $\mu$ and  $C\left(\frac{x}{y}, \frac{\mu}{yP_3}\right)$ is the matching kernel, evaluated so far to one-loop level in perturbation theory. To observe convergence to light-cone PDFs, it is not known \textit{a priori} how large the hadron boost must be in actual lattice computations. In fact, the largest achievable momentum remains limited by the exponential growth of the noise-to-signal ratio of the correlation functions and it cannot exceed the inverse of the lattice spacing, to avoid enhanced cutoff effects.

\section{Lattice details}

The lattice computation is performed using two ensembles of gauge field configurations generated by the Extended Twisted Mass Collaboration (ETMC).
The configurations were produced \cite{Alexandrou:2018egz} with twisted mass fermions \cite{Frezzotti:2000nk,Frezzotti:2003ni} including a clover term \cite{Sheikholeslami:1985ij} and Iwasaki gluons \cite{Iwasaki:1985we}.
For both ensembles, the lattice spacing is $a\approx0.093$ fm and they differ in the physical box size and the pion mass.
The details of simulation parameters are shown in Tab.~\ref{tab:ensemble}.

\begin{table}[h!]
	\centering
	\begin{tabular}{c|cc}
		\hline
		\hline
		\multicolumn{3}{c}{$\beta=1.726$    $c_{sw}=1.74$   $a=0.093$ fm} \\ \hline
		$24^3\times48$   & a$\mu$=0.0053 & $m_N$=1.21(2) GeV\\
	  L=2.2 fm & $m_\pi$=0.360 GeV &  $m_\Delta $=1.59(4) GeV\\ \hline
		$32^3\times64$ & a$\mu$=0.003 & $m_N$=1.08(3) GeV \\
		L=3.0 fm    & $m_\pi$=0.270 GeV &  $m_\Delta $=1.42(5) GeV\\ \hline
	\end{tabular}
	\caption{\label{tab:ensemble} Simulation parameters for the two ensembles of used gauge field configurations.}
\end{table}

\par
To evaluate the isovector $u-d$ unpolarized quasi-PDFs of the $\Delta^+$ baryon, we compute the matrix elements given by
\begin{equation}\label{equ:square}
h(P_3, z)=\left\langle h(P_3)\left|\overline{\psi}(0) \gamma^{0} W(0,z) \psi(z)\right| h(P_3)\right\rangle ,
\end{equation}
where $W(0,z)$ is a straight Wilson line with length typically varying from zero up to half of the lattice extension and $\gamma_0$ is the Dirac structure that offers a faster convergence to the unpolarized light-cone PDF \cite{Radyushkin:2016hsy}, preventing also mixing with other operators \cite{Constantinou:2017sej}. 
In order to extract the matrix element of Eq.~(\ref{equ:square}) in lattice QCD, we need the evaluation of two- and three-point correlation functions, given by
\begin{align}
C_{\sigma\rho}^{2 \mathrm{pt}}(\mathcal{P};\mathbf{P}; t_s) &=\mathcal{P}_{\alpha \beta} \sum_{\mathbf{x}_s} e^{-i \mathbf{P} \cdot \mathbf{x}_s}\left\langle 0\left|{\cal J}_{\sigma \alpha}(t_s,\mathbf{x}_s) \overline{{\cal J}}_{\rho\beta}(0,\mathbf{0})\right| 0\right\rangle , \label{eq:def_two_point}\\ 
C^{3 \mathrm{pt}}_{\sigma0\rho}\left(\tilde{\mathcal{P}};\mathbf{P}; t_{s}, t_{ins};z\right) &=\tilde{\mathcal{P}}_{\alpha \beta} \sum_{\mathbf{x}_s, \mathbf{x}_{ins}} e^{-i \mathbf{P} \cdot \mathbf{x}_s}\left\langle 0\left|{\cal J}_{\sigma\alpha}\left(t_s,\mathbf{x}_s\right) \mathcal{O}(t_{ins},\mathbf{x}_{ins}; z) \overline{{\cal J}}_{\rho\beta}(0,\mathbf{0})\right| 0\right\rangle ,
\label{eq:def_three-point}
\end{align}
where $x=(0,\mathbf{0})$ is the position at which the $\Delta^+$ is created (source), $x_s=(t_s,\mathbf{x}_s)$ the lattice point at which the $\Delta^+$ is annihilated  (sink) and $x_{ins}=(t_{ins},\mathbf{x}_{ins})$ indicates the lattice site at which the extended operator $\mathcal{O}$, of the form
\begin{equation}
\mathcal{O}(t_{ins},\mathbf{x}_{ins}; z)=\overline{\psi}(t_{ins},\mathbf{x}_{ins}) \gamma^0 W_z(\mathbf{x}_{ins}, \mathbf{z}_{ins}+z \hat{e}_z) \psi(t_{ins},\mathbf{x}_{ins}+z\hat{e}_z)
\end{equation}
couples to the quark fields. In Eqs.~(\ref{eq:def_two_point})-(\ref{eq:def_three-point}), $\mathcal{P}$ and $\tilde{\mathcal{P}}$ denote the parity projectors used for two- and three-point functions, respectively, here taken to be the same and equal to the positive parity projector $\Gamma=\frac{1+\gamma^{0}}{4}$.  For the interpolating field of $\Delta^+$, we use
\begin{equation}
{\cal J}_{\sigma\alpha}(x)=\frac{1}{\sqrt{3}} \epsilon^{a b c}\left[2\left(u^{a^{T}}(x) C \gamma_{\sigma} d^{b}(x)\right)u_{\alpha}^{c}(x)+\left(u^{a^{T}}(x) C \gamma_{\sigma} u^{b}(x)\right) d_{\alpha}^{c}(x) \right],
\label{eq:interpolating_field}
\end{equation}
with $C=\gamma^0\gamma^2$ being the charge conjugation matrix. The index $\sigma$ describes the polarization of the $\Delta^+$ baryon. This operator not only creates the $\Delta$ baryon
with spin $3/2$ from the QCD vacuum, but also spin-$1/2$ states which can be viewed as excited-state contamination. 
Previous works have shown that the contamination is negligible~\cite{Alexandrou:2008tn}.

The three-point function can also be written as
\begin{equation}
\begin{split}
C_{\sigma0\tau}^{3 \mathrm{pt}}\left(\tilde{\mathcal{P}};\mathbf{P} ; t_{s}, t_{ins};z\right)=\frac{M^{2}}{E^{2}}|Z|^{2} e^{-E t_{s}} h\left(P_3, z\right) \operatorname{tr}\left[\tilde{\mathcal{P}}_{\alpha \beta} \Lambda_{\sigma \sigma^{\prime}} \gamma^{0} \delta^{\sigma^{\prime} \tau^{\prime}} \Lambda_{\tau^{\prime} \tau}\right]+\cdots
\end{split}
\label{eq:three-point}
\end{equation}
where $M$ is the $\Delta$ mass and $E=\sqrt{M^2+\mathbf{P}^2}$ is its energy. $\Lambda$ is the summation of spinor in Euclidean space,
\begin{equation}
\begin{split}
\Lambda_{\sigma \tau}&=\sum_{s=-\frac{3}{2}}^{\frac{3}{2}} u_{\sigma}^{\Delta}(p, s) \overline{u}_{\tau}^{\Delta}(p, s)\\
&=-\frac{-i  p+M}{2 M}\left(\delta_{\sigma \tau}-\frac{\gamma_{\sigma} \gamma_{\tau}}{3}-\frac{2 p_{\sigma} p_{\tau}}{3 M^{2}}-i \frac{p_{\sigma} \gamma_{\tau}-p_{\tau} \gamma_{\sigma}}{3 M}\right),
\end{split}
\label{eq:Lambda}
\end{equation}
where $u_{\sigma}^{\Delta}(p, s)$ and $\overline{u}_{\tau}^{\Delta}(p, s)$ are the spinors for the $\Delta$ field and $p=(iE,\mathbf{P})$ is the Euclidean four-momentum.

To extract the desired matrix elements of Eq.~(\ref{equ:square}), we form ratios of three- over two-point functions, and seek for the region where the ratios are independent of the insertion time of the operator. For sufficiently large source-sink time separation and insertion time relative to the source, the plateau average identifies the matrix elements on the ground state. This approach, often known as \textit{plateau method} and used throughout this work, allows to extract the matrix elements as
\begin{equation}
h\left(\mathbf{P}, z\right)\stackrel{ 1\ll t_{ins}\ll t_s}{=}\frac{\sum_{\sigma=1}^{3} C_{\sigma 0 \sigma}^{3pt}\left(\tilde{\mathcal{P}};\mathbf{P}; t_s,t_{ins};z\right)}{ \sum_{\sigma=1}^{3} C^{2pt}_{\sigma \sigma}(\mathcal{P};\mathbf{P};t_s)}\; .
\label{eq:matrixelement}
\end{equation}

To increase the overlap of the interpolating field of Eq.~(\ref{eq:interpolating_field}) with the ground state of the boosted $\Delta^+$ baryon, we apply the momentum smearing technique~\cite{Bali:2016lva} on the quark fields. Our implementation of the smearing function reads
 \begin{equation}
 S_{mom}=\frac{1}{1+6\alpha} \Big( \psi(x)+\alpha \sum_j U_j(x) e^{-i\xi P \cdot \hat{j}} \psi(x+\hat{j}) \Big),
 \end{equation} 
where $\alpha$ is the Gaussian smearing~\cite{Gusken:1989qx,Alexandrou:1992ti} parameter, $U_j$ the APE-smeared~\cite{Albanese:1987ds} gauge links along the direction $j$ of the momentum and $\xi$ is a parameter that needs to be optimized to minimize the noise-to-signal ratio on the correlation functions. The parameters employed for the Gaussian and APE-smearing have been, in turn, tuned to improve the quality of the signal and obtain a smooth smearing function.  

\begin{figure}[h]
 	\centering
    \includegraphics[width=0.5\textwidth]{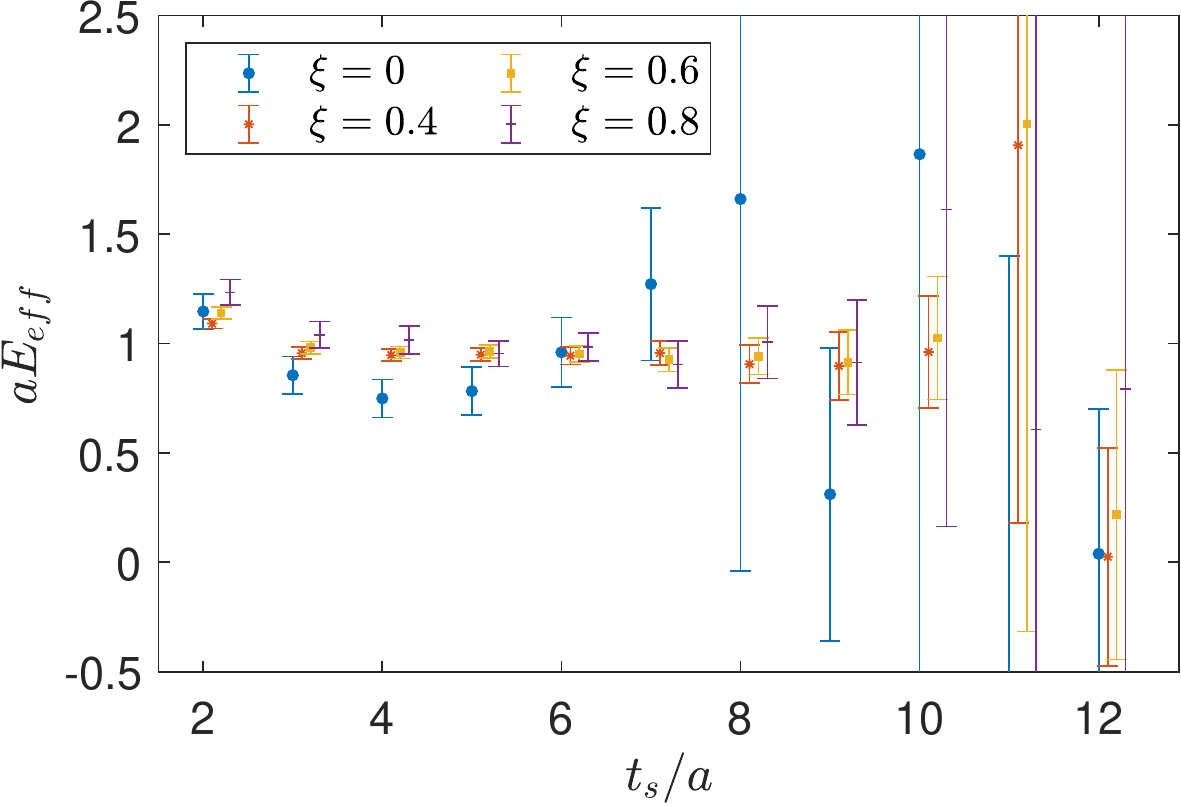}
    \hspace{5mm}
 	\caption{Effective energies of the $\Delta$ baryon boosted to $P_3=6\pi/L$ (270 MeV pion mass ensemble) for different values of $\xi$. }
 	\label{fig:32mom003smear}
 \end{figure}

 In Fig.~\ref{fig:32mom003smear}, we show the effective energies from momentum-smeared two-point functions for different values of $\xi$, for the 270 MeV pion mass ensemble at the largest of our boosts, $P_3=6\pi/L$.
 The statistical errors are reduced by a large $t_s$-dependent factor with respect to $\xi=0$ when using any of the three non-zero values of $\xi$ shown in the plot.
 The values of $\xi=0.4$ and $\xi=0.6$ lead to smallest and comparable errors at source-sink separations relevant to the evaluation of matrix elements.

\section{Renormalization Procedure}
\label{sec:renorm} 

To renormalize the bare matrix elements discussed in the previous sections, we use the non-perturbative procedure developed in Ref.~\cite{Alexandrou:2017huk}. The latter uses the renormalization pattern found in the perturbative calculation of non-local fermion operators for a Wilson-type lattice formulation~\cite{Constantinou:2017sej}. The proposed prescription eliminates, at once, all divergences, including the power-law divergence due to the presence of the Wilson line for specific regularization schemes~\cite{Dotsenko:1979wR,Brandt:1981kf}, including the lattice. In general, such a complete renormalization procedure is necessary before taking the continuum limit and comparing it with phenomenological and experimental data, if available.

Here, we briefly outline the renormalization program, which follows the procedure employed for nucleon quasi-PDFs, described in detail in Ref.~\cite{Alexandrou:2019lfo}. The non-perturbative procedure relies on an RI$^\prime$-scheme~\cite{Martinelli:1994ty}, which is generalized for operators including a Wilson line. The renormalization function for the unpolarized PDFs, $Z_{\rm V0}$, is extracted using the temporal direction for the Dirac matrix, as we do for the matrix elements. This choice is necessary to avoid the mixing between operators for the unpolarized case in which the Dirac matrix is in the same direction as the Wilson line~\cite{Constantinou:2010gr}. In the absence of mixing, the RI$^\prime$-scheme condition is given by
\begin{equation}
\label{renorm}
\frac{Z^{\RI}_{\rm V0}(z,\mu_0,m_\pi)}{Z^{\RI}_q(\mu_0,m_\pi)}\frac{1}{12} {\rm Tr} \left[{\cal V}(z,p,m_\pi) \left({\cal V}^{\rm Born}(z,p)\right)^{-1}\right] \Bigr|_{p^2{=}\mu_0^2} {=} 1\, ,
\end{equation}
where $Z_q$ is the renormalization function of the quark field, obtained from
\begin{equation}
Z^{\RI}_q(\mu_0,m_\pi) \frac{1}{12} {\rm Tr} \left[(S(p,m_\pi))^{-1}\, S^{\rm Born}(p)\right] \Bigr|_{p^2=\mu_0^2}  = 1 \,. \hspace*{1.4cm}
\end{equation}
Both $Z_{\rm V0}$ and $Z_q$ are scheme and scale dependent\footnote{Note that $Z_{\rm V0}$ for $z=0$ reduces to the renormalization function of the local vector current, which is scheme and scale independent.}, and are expected to have some dependence on the pion mass. ${\cal V}$ is the amputated vertex function of the operator and $S$ the fermion propagator, while ${\cal V}^{{\rm Born}}$ and $S^{{\rm Born}}$ are the corresponding tree-level values. The condition of Eq.~(\ref{renorm}) is applied for each value of $z$. The RI$^\prime$ renormalization scale, $\mu_0$, is chosen to have equal spatial directions, 
\begin{equation}
a\mu_0=2 \pi\left(\frac{n_t+1/2}{L_t},\frac{n}{L_s},\frac{n}{L_s},\frac{n}{L_s}\right)\,,
\end{equation}
where $L_s$ ($L_t$) is the spatial (temporal) extent of the lattice. We choose the integers $n_t$ and $n$ such that the ratio $P4{\equiv}{\sum_i \mu_i^4}/{(\sum_i \mu_i^2)^2}$ is close to or at its minimum value of 1/4. One of the benefits of this choice is the reduction of the Lorentz non-invariant contributions~\cite{Constantinou:2010gr,Alexandrou:2015sea}. In our calculation, we use the combinations of $n_t \in [3-9]$ and $n \in [2-4]$ that satisfy $P4{\equiv}{\sum_i \mu_i^4}/{(\sum_i \mu_i^2)^2}<0.28$ and $(a\mu_0)^2 \in [1-5]$ (a total of 12 scales). 

As in our previous work for the nucleon quasi-PDFs, we employ the momentum source method~\cite{Gockeler:1998ye,Alexandrou:2015sea} that offers high statistical accuracy. We produce the vertex functions on five $N_f=4$ ensembles given in Tab.~\ref{Table:Z_ensembles}, which differ only in the pion mass. The gauge configurations used here were produced for renormalization functions, and correspond to the same $\beta$ value as the $N_f=2+1+1$ ensemble used for the matrix elements. The necessity of degenerate up, down, strange and charm quark is to allow one to take a proper chiral extrapolation.
\begin{table}[h]
\begin{center}
\renewcommand{\arraystretch}{1.5}
\renewcommand{\tabcolsep}{5.5pt}
\begin{tabular}{ccc}
\hline\hline 
$\beta=1.726$, & $c_{\rm SW} = 1.74$, & $a=0.093$~fm \\
\hline\hline\\[-3ex]
{$24^3\times 48$}  & {$\,\,a\mu = 0.0060$}  & $\,\,m_\pi = 357.84$~MeV     \\
\hline
{$24^3\times 48$}  & $\,\,a\mu = 0.0080$     & $\,\,m_\pi = 408.11$~MeV     \\
\hline
{$24^3\times 48$}  & $\,\,a\mu = 0.0100$    & $\,\,m_\pi = 453.48$~MeV    \\
\hline
{$24^3\times 48$}  & $\,\,a\mu = 0.0115$    & $\,\,m_\pi = 488.41$~MeV    \\
\hline
{$24^3\times 48$}  & $\,\,a\mu = 0.0130$    & $\,\,m_\pi = 518.02$~MeV    \\
\hline\hline
\end{tabular}
\vspace*{-0.25cm}
\begin{center}
\caption{\small{Parameters of the $N_f=4$ ensembles used for the calculation and chiral extrapolation of $Z_{\rm V0}$}.}
\label{Table:Z_ensembles}
\end{center}
\end{center}
\vspace*{-0.2cm}
\end{table} 

Using the above ensembles we apply a chiral extrapolation on the real and imaginary parts, of the form
\begin{equation}
\label{eq:Zchiral_fit}
Z^{\RI}_{\rm V0}(z,\mu_0,m_\pi) = {Z}^{\RI}_{{\rm V0},0}(z,\mu_0) + m_\pi^2 \,{Z}^{\RI}_{{\rm V0},1}(z,\mu_0) \,,
\end{equation}
where $ {Z}^{\RI}_{{\rm V0},0}(z,\mu_0) $ is the desired chiral-limit value, which is calculated on each RI$^\prime$ renormalization scale. As in Ref.~\cite{Alexandrou:2019lfo}, we find that $Z_{\rm V0}$ has negligible dependence on the pion mass for small values of $z$ (${Z}^{\RI}_{{\rm V0},1}(z,\mu_0)\sim 0$), which becomes linear in $m_\pi^2$ as $z$ increases (typically $z/a\ge5$).

The conversion to the $\overline{\rm MS}$ and evolution to 2 GeV follows the chiral extrapolation. This procedure relies on the results obtained in one-loop perturbation theory using dimensional regularization~\cite{Constantinou:2017sej}. Higher-loop results are currently not available because of the difficulty of isolating the IR and UV divergences. A discussion on the truncation effects due to the conversion factor can be found in Ref.~\cite{Alexandrou:2019lfo}. The $\overline{\rm MS}$-scheme estimates are then converted to the modified $\msbar$ scheme ($\mmsbar$), for which the corresponding matching formula satisfies particle number conservation~\cite{Alexandrou:2019lfo}.

\begin{figure}[h!]
\includegraphics[scale=.64]{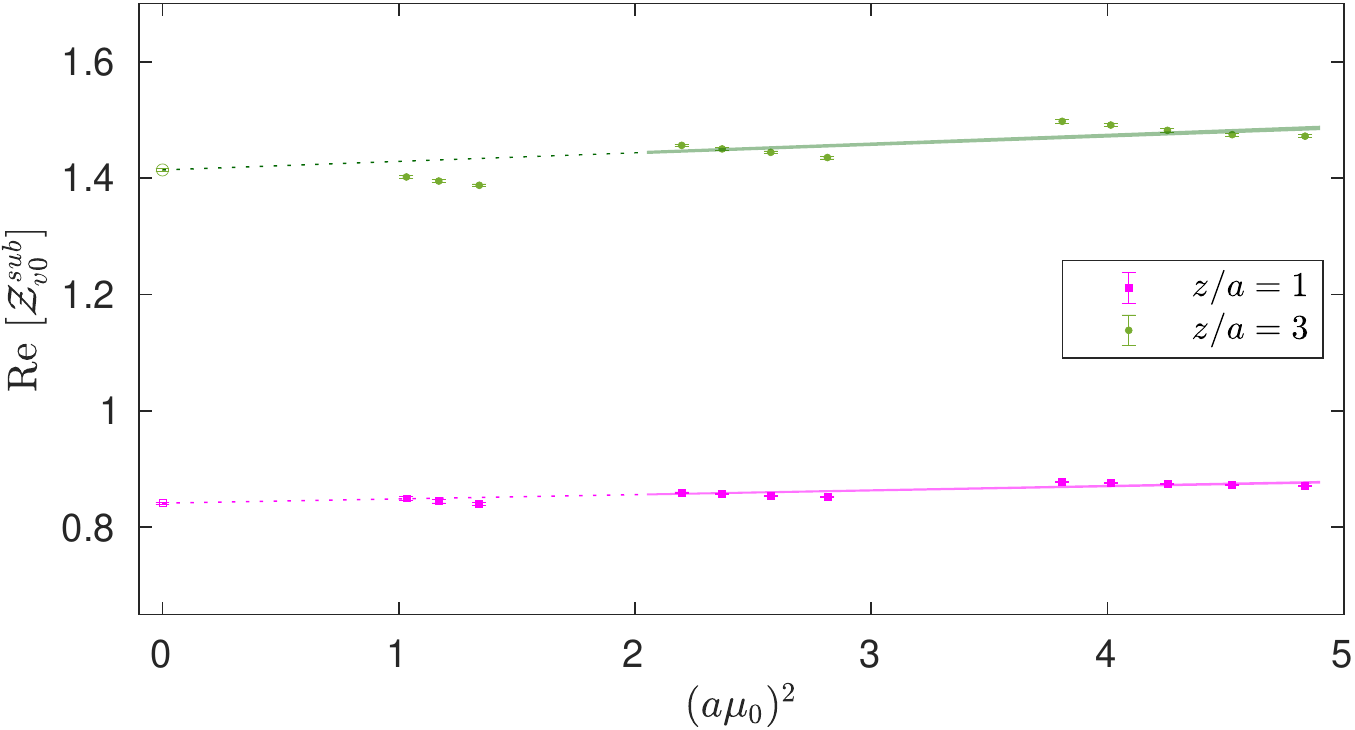}
\includegraphics[scale=.64]{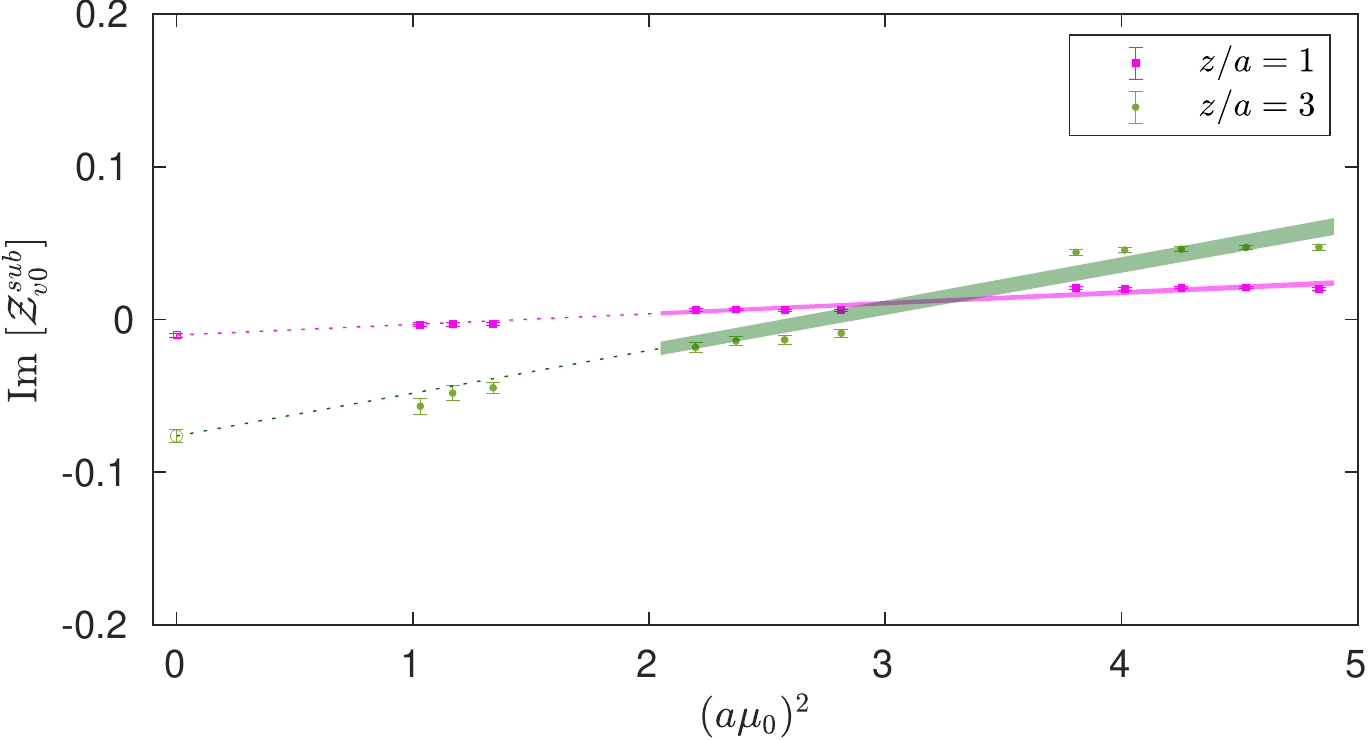}
\vspace*{-0.15cm}
\caption{Real (left) and imaginary (right) part of ${Z}^{\MMSb}_{\rm V0}(z,\bar\mu,\mu_0)$ as a function of the initial RI$^\prime$ scale. The upper (lower) panel corresponds to $z/a=1$ ($z/a=3$). The dashed line corresponds to the fit of Eq.~(\ref{eq:Z_mu_fit}), and the open symbols are the extrapolated values  ${\cal Z}^{\MMSb}_{{\rm V0}}$.}
\label{fig:Zfactors_RI_scale_dependence}
\end{figure}

The final results for $Z_{\rm V0}$ in the $\mmsbar$-scheme evolved to 2 GeV have residual dependence on the initial renormalization scale $\mu_0$. To eliminate the unwanted dependence on $\mu_0$, we extract $Z_{\rm V0}$ for all values of $(a\mu)^2 \in [1-5]$ and perform a fit of the form
\begin{equation}
\label{eq:Z_mu_fit}
{Z}^{\rm M\MSb}_{\rm V0}(z,\bar\mu,\mu_0) = {\cal Z}^{\rm M\MSb}_{{\rm V0}}(z,\bar\mu) + (a\,\mu_0)^2 \,{\zeta}^{\rm M\MSb}_{{\rm V0}}(z,\bar\mu) \,.
\end{equation}
The desired quantity is the fit parameter ${\cal Z}^{\rm M\MSb}_{{\rm V0}}(z,\bar\mu)$.
In Fig.~\ref{fig:Zfactors_RI_scale_dependence}, we show ${\cal Z}^{\MMSb}_{{\rm V0}}(z,\bar\mu)$ as a function of the initial scale $(a\,\mu_0)^2$. For simplicity, we choose representative values of the length of the Wilson line, namely $z/a=1$ and $z/a=3$. We find that the imaginary part  has a stronger dependence on the $\mu_0$ value, similarly to our previous work for the nucleon quasi-PDFs. Such dependence has a small effect on the renormalized matrix elements, as the values of the imaginary part are sub-leading compared to the real part. Note that sizable discretization effects are found in several calculations of renormalization functions, even for local operators. To suppress this effect, we employ a procedure of subtracting finite-$a$ contributions using one-loop perturbation theory, which is very successful for local operators~\cite{Constantinou:2010gr,Constantinou:2014fka,Alexandrou:2015sea}. Here we partly improve our estimates by removing the artifacts from $Z_q$~\cite{Alexandrou:2015sea}, using the procedure outlined in Ref.~\cite{Alexandrou:2019lfo}. 

One important aspect of the fit given in Eq.~(\ref{eq:Z_mu_fit}) is the choice of the optimal fit region for $(a\mu_0)^2$. The latter has to be chosen to suppress non-perturbative effects $(a\mu_0)^2 > 1-2$, but should not be very large, for finite-$a$ effects to be small.
We test various intervals, and we choose as final results the values obtained from $(a\,\mu_0)^2 \in [2-5]$ for both the real and imaginary part. 

\begin{figure}[h!]
 	\centering
 	\subfigure{\includegraphics[width=0.4\textwidth]{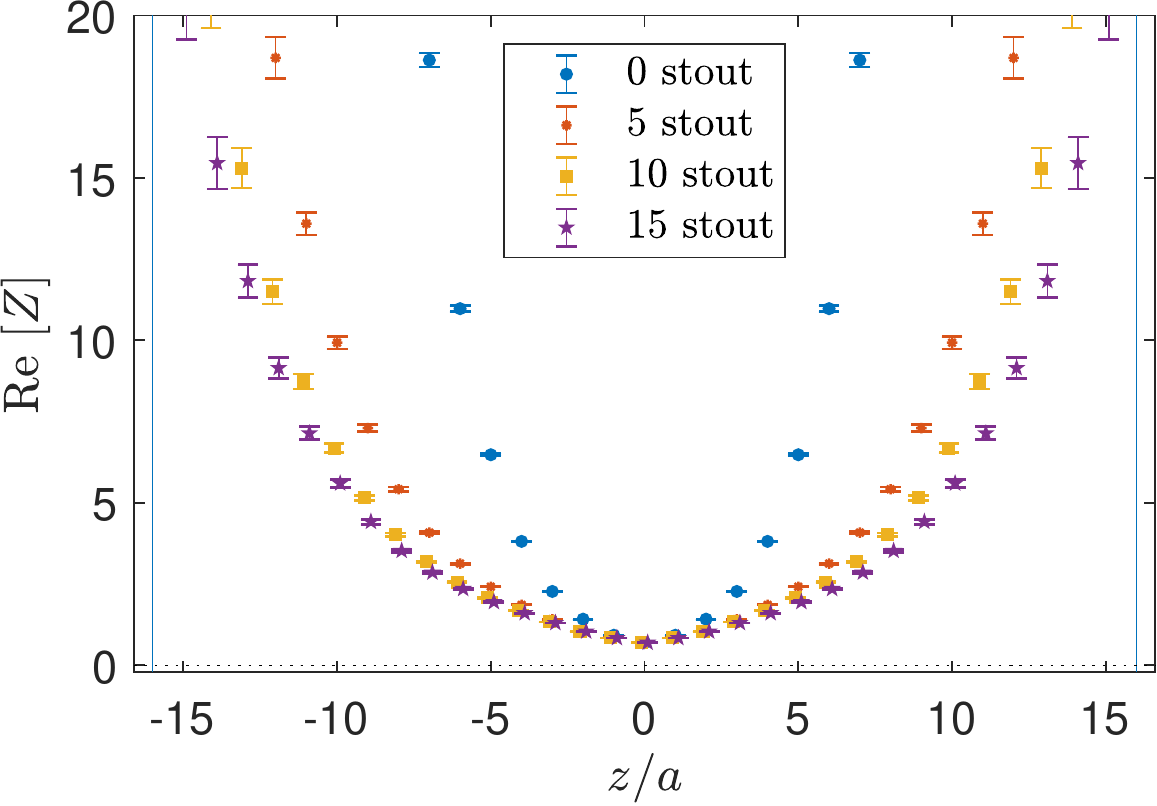}}\hspace{5mm}
 	\subfigure{\includegraphics[width=0.4\textwidth]{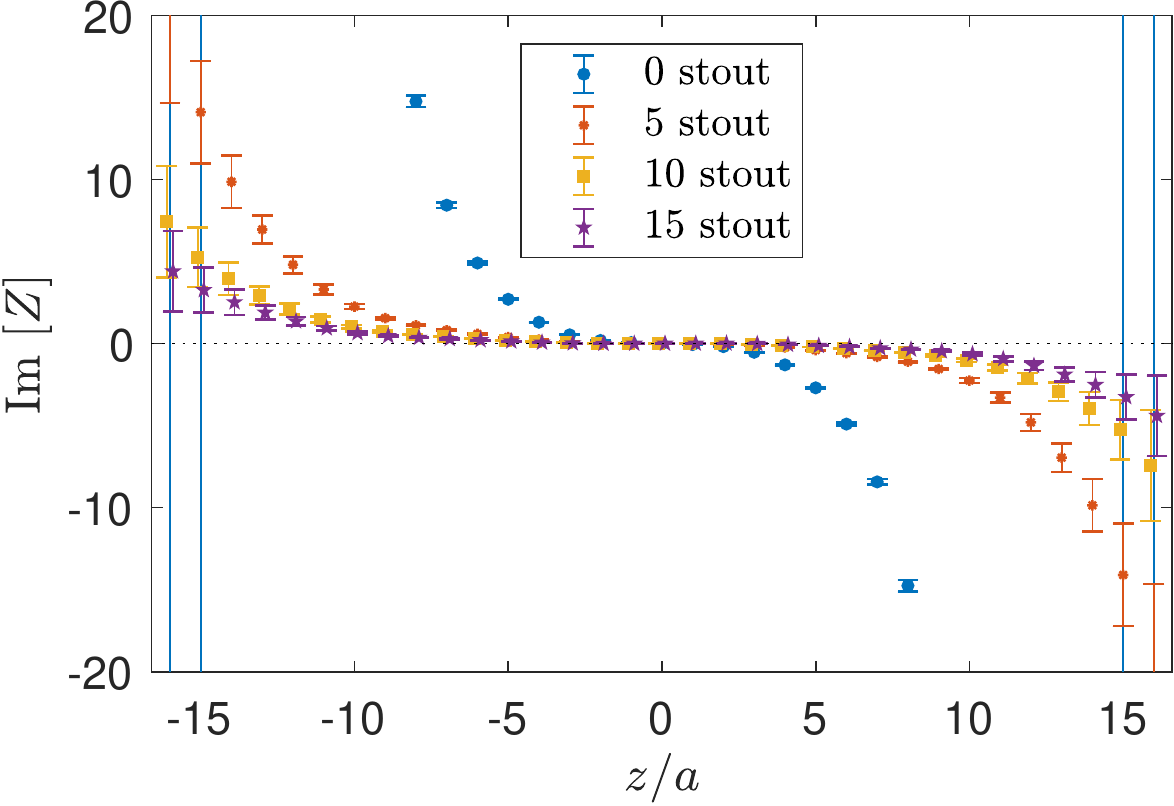}}
 	\caption{\small{Final values of $Z_{\rm V0}$ in the $\mmsbar$ scheme after chiral and $(a\,\mu_0)^2$ extrapolation. Real and imaginary parts are shown in the left and right panels, respectively.}}
 	\label{fig:Z_final}
 \end{figure}

The final values for $Z_{\rm V0}$ corresponding to 0, 5, 10 and 15 stout smearing steps are shown in Fig.~\ref{fig:Z_final}. We find that the statistical uncertainties of $Z_{\rm V0}$ are much smaller as compared to those of the matrix elements, due to the use of the momentum source method. One of the features of the renormalization functions of non-local operators is their rapid increase as the length of the Wilson line increases. This is due to the power-law divergence of the Wilson line. Finally, these values are used to obtain the renormalized matrix elements from the complex multiplication 
\begin{equation}
h^{\MMSb}_\Gamma(P_3,z, \bar\mu) =  h^{bare}_\Gamma(P_3,z) \cdot {\cal Z}^{\MMSb}_{\Gamma}(z,\bar\mu)\,.
\end{equation}

\section{Results}
In this section, we present our results using the two ensembles of gauge field configurations. 
The first ensemble is smaller in volume ($24^3\times48$) and reproduces a mass for the $\Delta$ equal to $m_\Delta=1.59(4)$ GeV, i.e.\ around 30\% heavier than its physical value. The masses for the nucleon and pion are $m_N=1.21(2)$ GeV and $m_\pi=0.36(1)$ GeV, respectively. Therefore, the difference between the mass of $\Delta$ and the sum of the nucleon and pion mass is very small. Within the full uncertainties of the baryon masses extraction, it can correspond to either side of the decay threshold. However, the decay is strongly suppressed and it is plausible to treat such unphysical $\Delta$ as stable. The second ensemble has a volume $32^3\times64$, and corresponds to $m_\Delta=1.42(5)$ GeV and $m_N=1.08(3)$ GeV, which is approximately 15\% heavier than in nature. The pion mass is $m_\pi=0.27(2)$ GeV, which is smaller than the mass difference 
between the two baryons. Therefore, $\Delta$ decay can occur.
However, this mass difference is close to the decay threshold.
Thus, similarly to the case of our smaller-volume ensemble, the phase space allowed for nucleon-pion decay is strongly suppressed and we assume the $\Delta$ baryon can still be treated as stable. This is a simplification that needs to be further investigated in the future, but its proper analysis will require a highly non-trivial extension of the quasi-distribution framework to the case of unstable hadrons. Nevertheless, given the strong suppression of the decay at our simulation parameters, it is plausible to assume the ensuing effects of the decay are subleading with respect to our current statistical and systematic precision.

Given the exploratory nature of this work for the $\Delta$ baryon, we first test the feasibility of extracting the relevant matrix elements using the ensemble with $m_\pi=360$ MeV and a momentum boost of 0.54 GeV. This is presented in Sec.~\ref{subsec360MeV}. Following the encouraging conclusions from this ensemble, we focus on the larger-volume ensemble at  $m_\pi=270$ MeV, to provide the main results of this work. In Sec.~\ref{subsec270MeV}, we show the bare and renormalized matrix elements, as well as the matching to light-cone PDFs for each momentum.

\subsection{Small-volume ensemble at $m_\pi=360$ MeV}
\label{subsec360MeV}
The exploratory results we present here are obtained using the $24^3 \times 48$ ensemble from Tab.~\ref{tab:ensemble}, with lattice spacing $a=0.093$~fm, physical extent around 2.2 fm and a pion mass of approx.\ 360 MeV. Our calculation was performed only at the smallest non-zero hadron momentum, $P_3=2\pi/L\approx0.54$~GeV.
The main aim was to observe the quality of the signal and to check the signal decay when increasing the source-sink separation, $t_s$. However, we postpone a more thorough investigation of excited states effects to a follow-up paper dedicated to  systematic uncertainties of the $\Delta$ PDFs.
We used four values of the source-sink separation, $t_s=\{9a, 10a, 11a, 12a\}\approx\{0.84, 0.93, 1.02, 1.12\}$~fm, with 886 measurements for the lower two $t_s$ values and 768 for the larger two.
No smearing was applied to the Wilson line entering the inserted operator.

\begin{figure}[h!]
 	\centering
 	\subfigure{\includegraphics[width=0.48\textwidth]{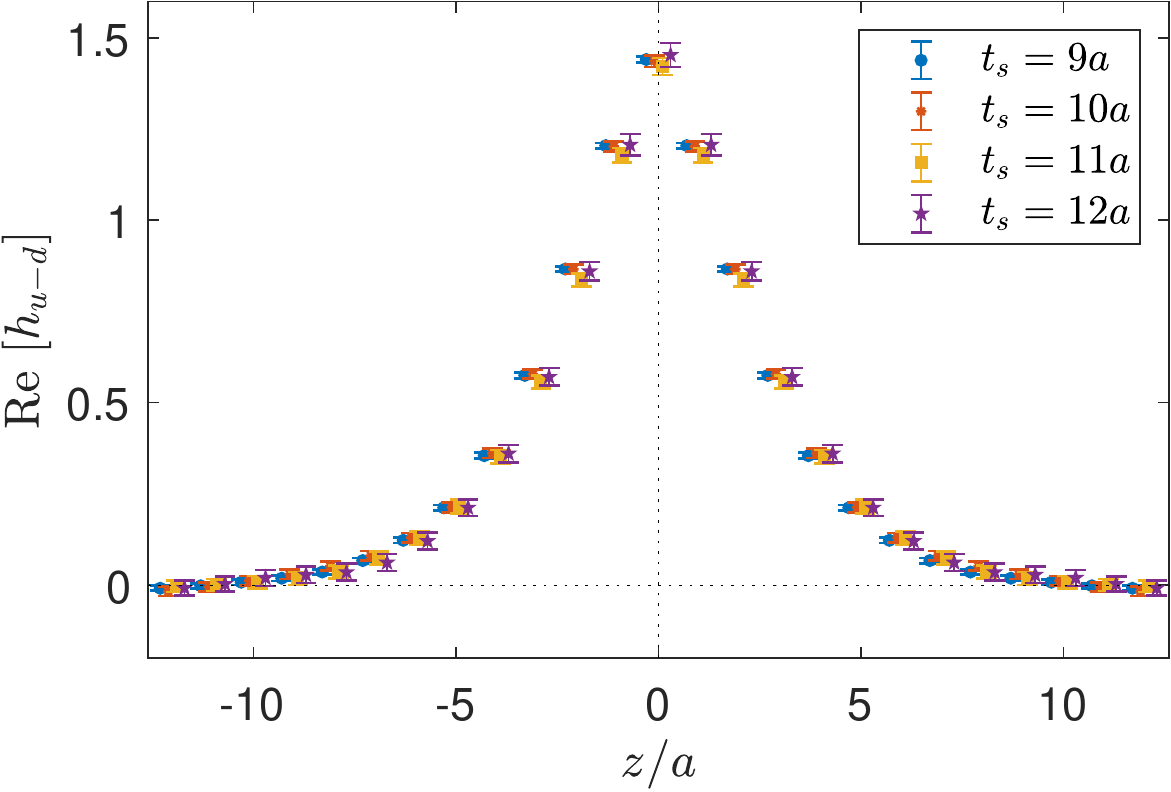}}
 	\subfigure{\includegraphics[width=0.48\textwidth]{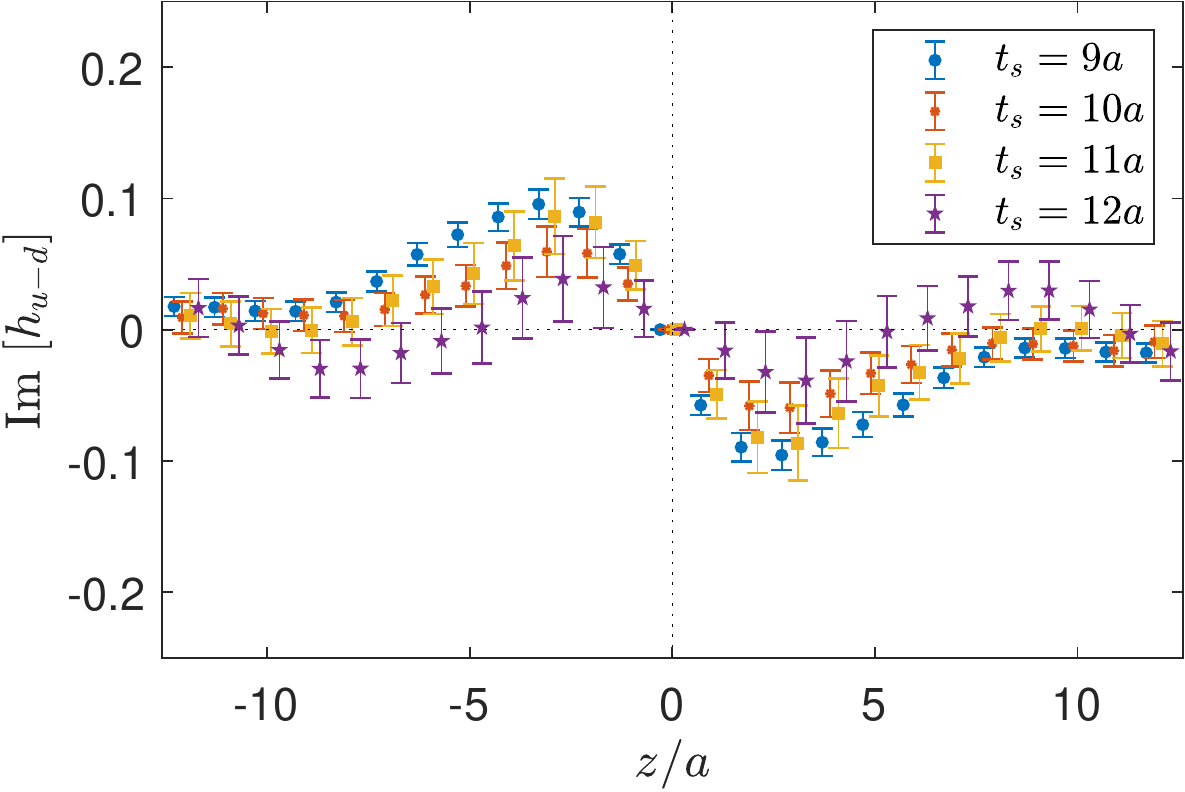}}
 	\caption{The real (left) and imaginary (right) part of the bare matrix elements defined in Eq.~(\ref{eq:matrixelement}) as a function of $z$ for four different source-sink separations.} 
 	\label{differnet time separation}
 \end{figure}

The real and imaginary parts of the extracted matrix elements at the four values of the source-sink separation are compared in Fig.~\ref{differnet time separation}.
We observe that the errors increases by a factor between 1.3-1.8 when the $t_s$ is increased by one lattice spacing. Comparison of these results may be used for investigation of excited-states contamination, within the statistical uncertainties. 
We find that the real part is compatible for all $t_s$ value at all lengths of the Wilson line, suggesting small effects from excited states.
For the imaginary part, we observe convergence for the three largest source-sink separations.
The data at the lowest separation tend to have a systematically larger magnitude, suggesting possible excited states contamination.
However, the imaginary part is small for the parameters of this calculation and hence, the precision is insufficient to draw meaningful conclusions.

\subsection{Larger volume ensemble at $m_\pi = 270$ MeV}
\label{subsec270MeV}
The main part of the work concentrates on the $32^3 \times 64$ ensemble 
from Tab.~\ref{tab:ensemble}, at the same lattice spacing as the $24^3 \times 48$ ensemble, but with a larger volume with physical extent $\sim$3 fm and a lighter pion mass of approx.\ 270 MeV. As discussed in Ref.~\cite{Ethier:2018efr}, the sea antiquark asymmetry is expected to grow as the decay channel is approached, 
and how far we are from the decay channel will depend on the 
momentum given to the $\Delta$ baryon. Our results for the dependence of the asymmetry on the injected momentum will be presented at the end of this section.
We use a source-sink separation of $t_s=10a$, which, based on the previous subsection, is sufficient to suppress excited states within the current statistical uncertainties.
However, we emphasize that further studies need to be performed for excited states contamination, since for the present ensemble we use larger momenta and the pion mass is smaller, both of which lead to enhanced excited states effects.
Nevertheless, it is beyond the scope of the present study.
The momenta we use are the first three discrete lattice momenta at this volume, $P_3=\{2\pi/L,4\pi/L,6\pi/L\}$, which correspond to approx.\ $\{0.42,0.83,1.25\}$ GeV. 
We aim for similar precision for all three momenta and thus, we increase statistics when the hadron boost is increased, namely we perform 906, 8784 and 42660 measurements, respectively.

\begin{figure}[htbp]
 	\centering
 	\subfigure{\includegraphics[width=0.48\textwidth]{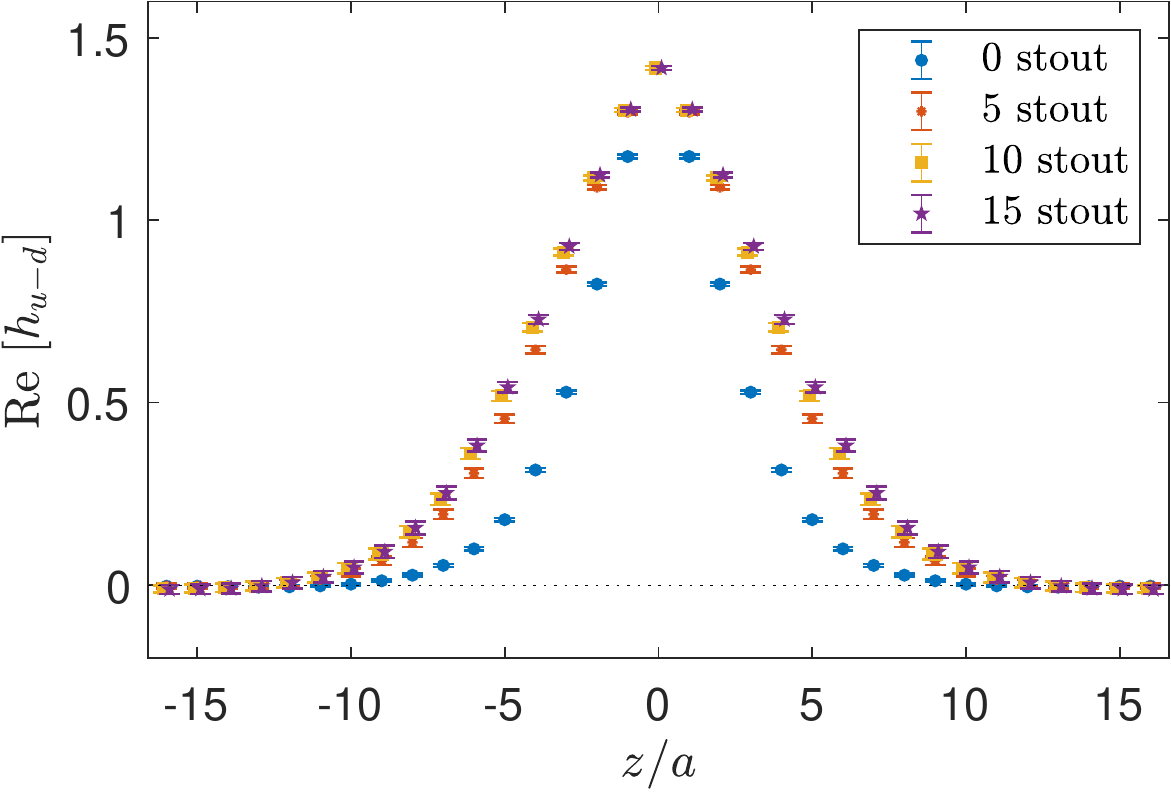}}\hspace{5mm}
 	\subfigure{\includegraphics[width=0.48\textwidth]{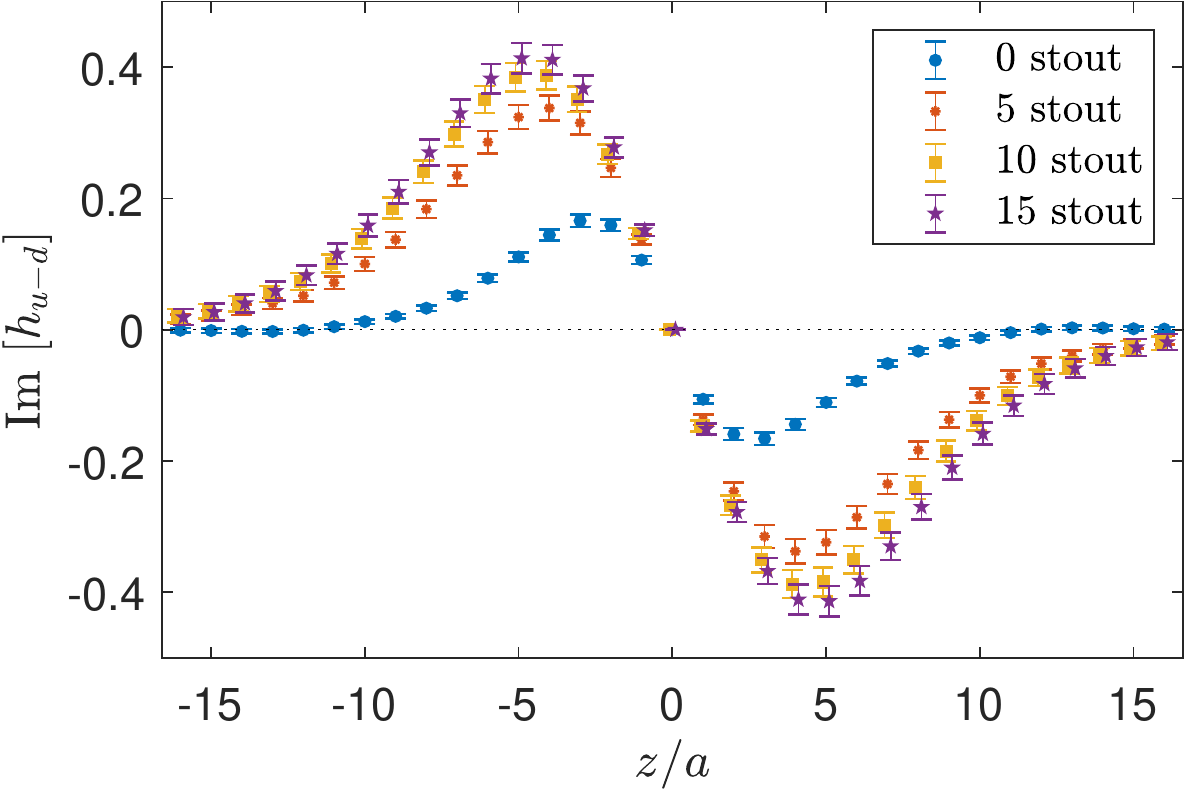}}
 	\caption{The real (left) and imaginary (right) part of the bare matrix elements for 0, 5, 10 and 15 steps of stout smearing steps. The hadron boost is $P_3=6\pi/L$.}
 	\label{matrix element}
 \end{figure}

\begin{figure}[h]
 	\centering
 	\subfigure{\includegraphics[width=0.48\textwidth]{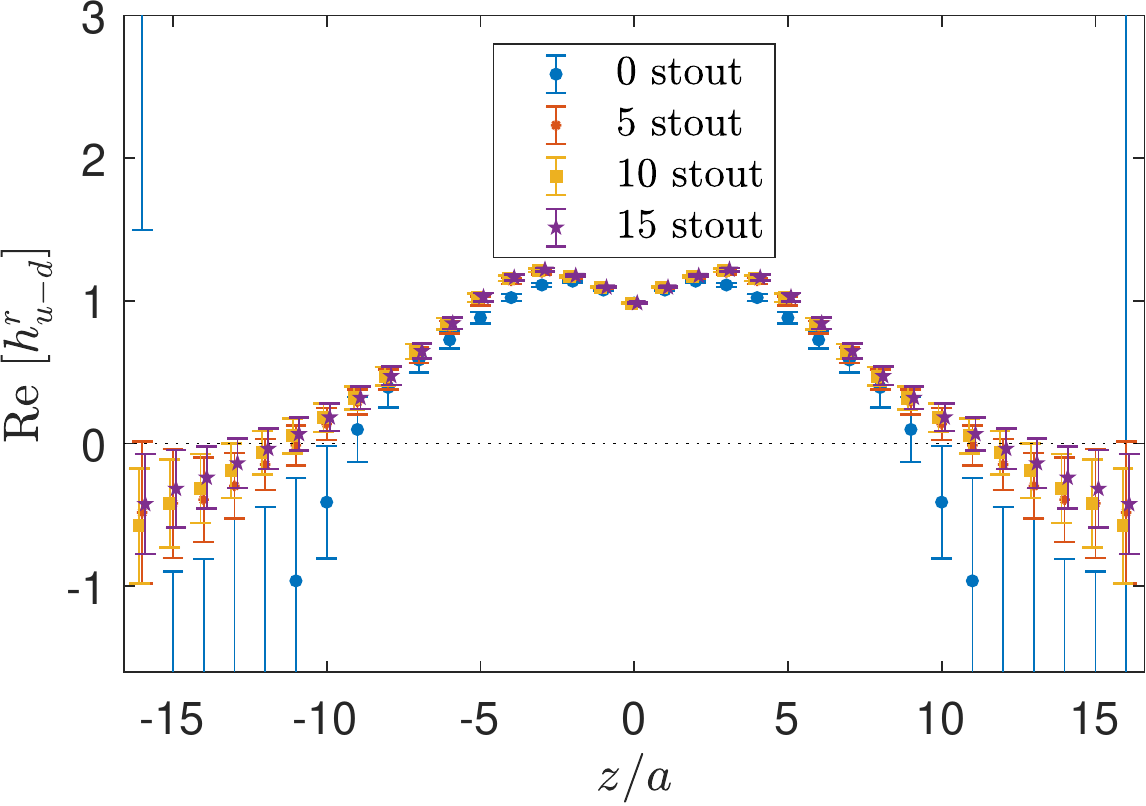}}\hspace{5mm}
 	\subfigure{\includegraphics[width=0.48\textwidth]{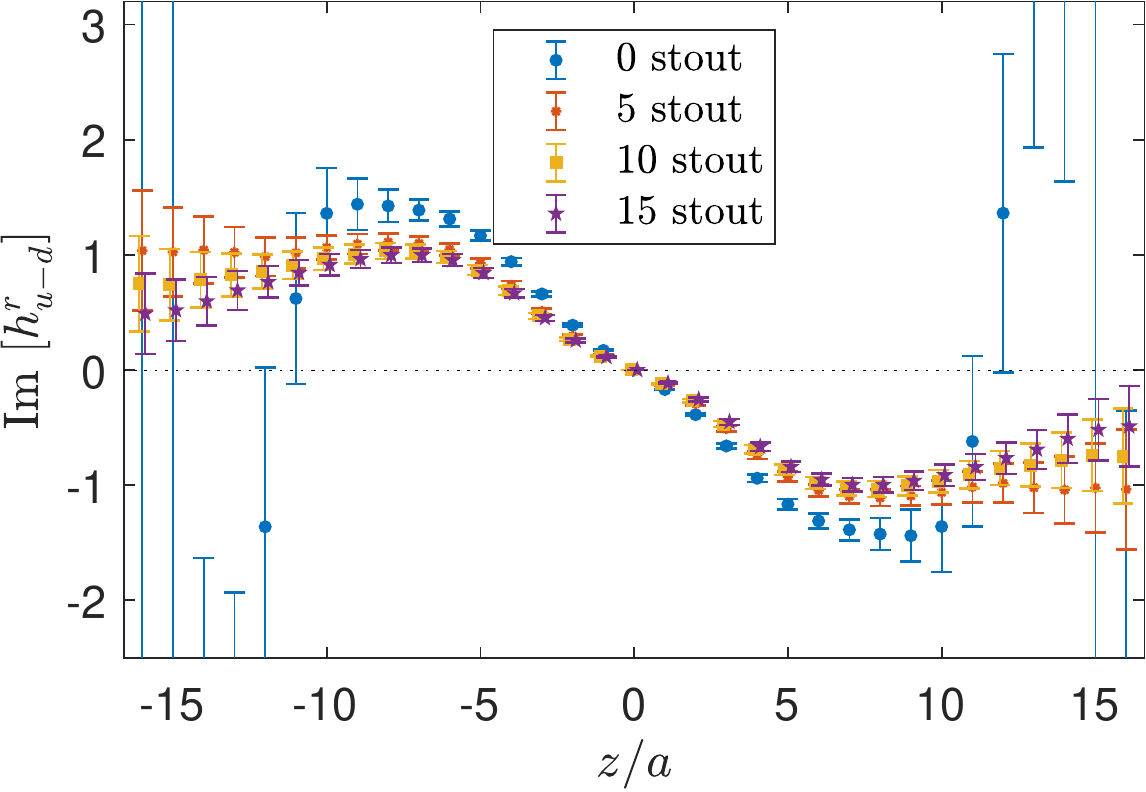}}
 	\caption{The real (left) and imaginary (right) part of the renormalized matrix elements for 0, 5, 10 and 15 steps of stout smearing steps. The hadron boost is $P_3=6\pi/L$.}
 	\label{renormalization matrix}
 \end{figure}

It is interesting to compare the quality of the signal for the $\Delta$ and the nucleon. We performed such a test at momentum $P_3=6\pi/L$, and we found that the $\Delta$ is around $30\%$ more noisy than the nucleon. This makes the study of systematic uncertainties more challenging for the $\Delta$ baryon.
Next, we investigate the effects of applying three-dimensional stout smearing~\cite{Morningstar:2003gk} to the gauge links that enter in the Wilson line of the operator. 
This reduces the power divergence in the matrix elements and makes the renormalization functions closer to their tree-level values.
Moreover, the renormalization functions are obtained with much better statistical precision when stout smearing is applied, see Sec.~\ref{sec:renorm}.
In Fig.~\ref{matrix element}, we show the bare matrix elements for momentum $6\pi/L$ with 0, 5, 10 and 15 stout smearing steps. 
As expected, the number of stout steps affects the values of bare matrix elements. 
However, the renormalized matrix elements must be independent of the level of stout smearing, since renormalized matrix elements are related to physical quantities.
We apply the renormalization procedure described in Sec.~\ref{sec:renorm} calculated for each stout step, and the renormalized matrix elements are compared in Fig.~\ref{renormalization matrix}.
We find that these matrix elements are consistent with one another for different number (nonzero) of stout smearing iterations, thus validating that the influence of the smearing procedure on the power divergence present in the bare matrix elements is correctly captured in the renormalization functions. We find small deviation in the case of stout=0, which is due to systematic uncertainties in the renormalization. In the remainder of the paper, we take results from 10 steps of stout smearing.

 
\begin{figure}[h]
 	\centering
 	\subfigure{\includegraphics[width=0.48\textwidth]{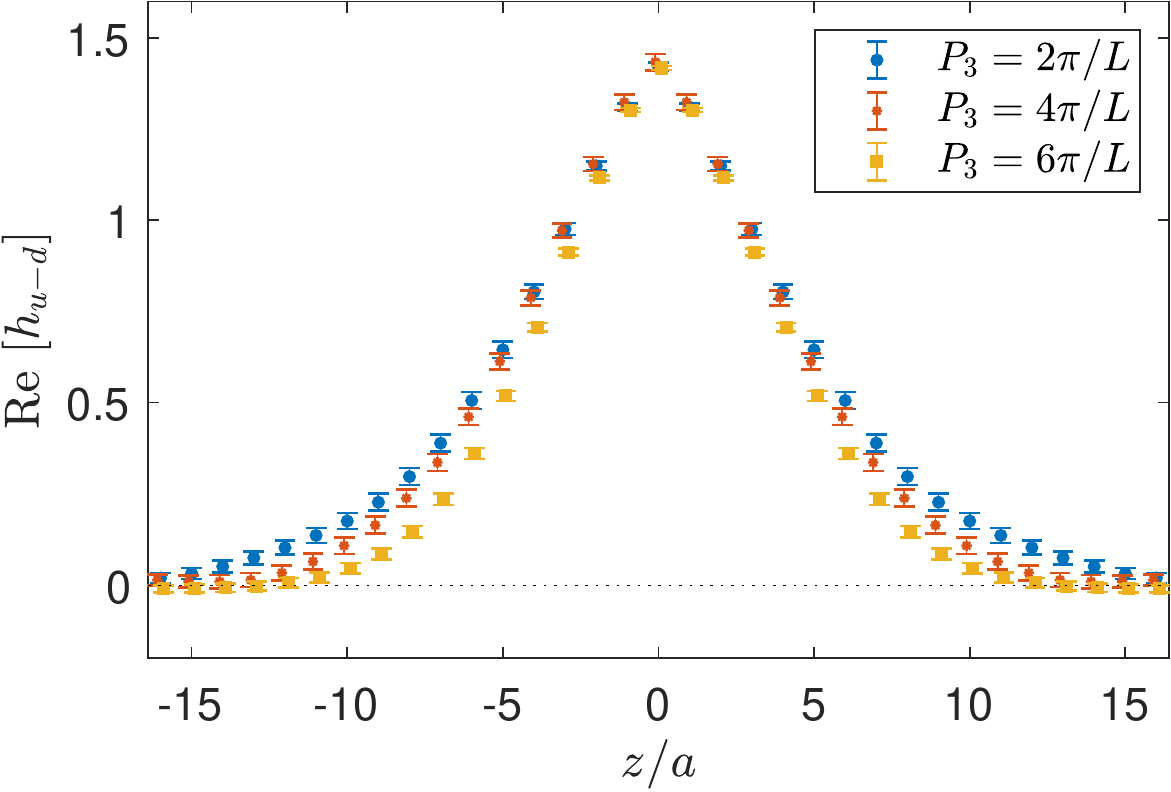}}\hspace{5mm}
 	\subfigure{\includegraphics[width=0.48\textwidth]{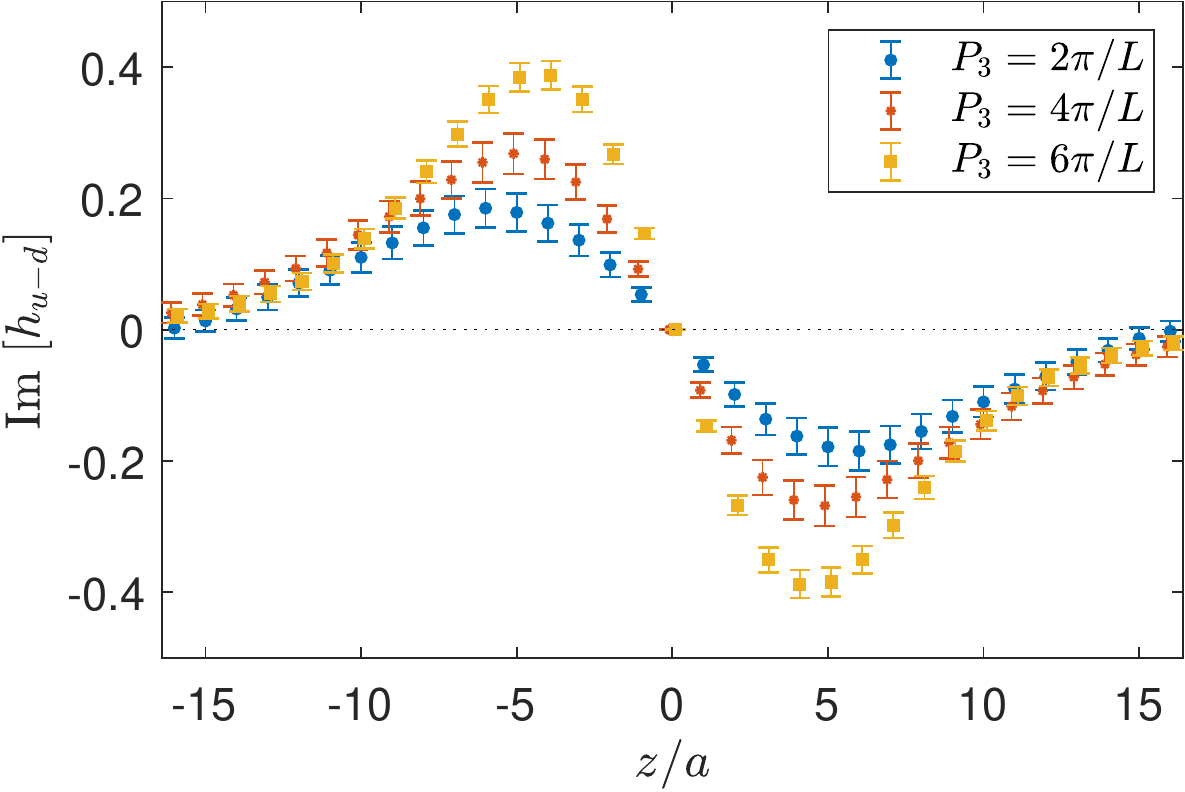}}
 	\caption{The real (left) and imaginary (right) part of bare matrix elements for the three lowest lattice momenta, $P_3=2\pi/L,4\pi/L,6\pi/L$, with 10 steps of stout smearing.}
 	\label{different momentum}
 \end{figure}

 Another interesting study is the influence of momentum boost on the matrix elements.
 Clearly, it is desirable to reach as large momentum of the hadron as possible.
 However, the signal-to-noise ratio decays exponentially when the momentum is increased and moreover, excited states contamination may become difficult to control at large boosts.
 In Fig.~\ref{different momentum}, we show bare matrix elements for the three lowest lattice momenta with 10 steps of stout smearing. 
We increase the number of measurements considerably for higher values of $P_3$, in order to keep the statistical errors similar for all three momenta.
 We achieve the smallest errors for the largest boost, which is due to an almost 50-fold increase of statistics with respect to the lowest momentum. 
 The influence of the boost on the matrix elements is consistent with expectations from nucleon studies -- larger hadron momentum implies faster decay of the matrix elements to zero in the real part and a more pronounced peak in the imaginary part.

\begin{figure}[h]
 	\centering
 	\subfigure{\includegraphics[width=0.48\textwidth]{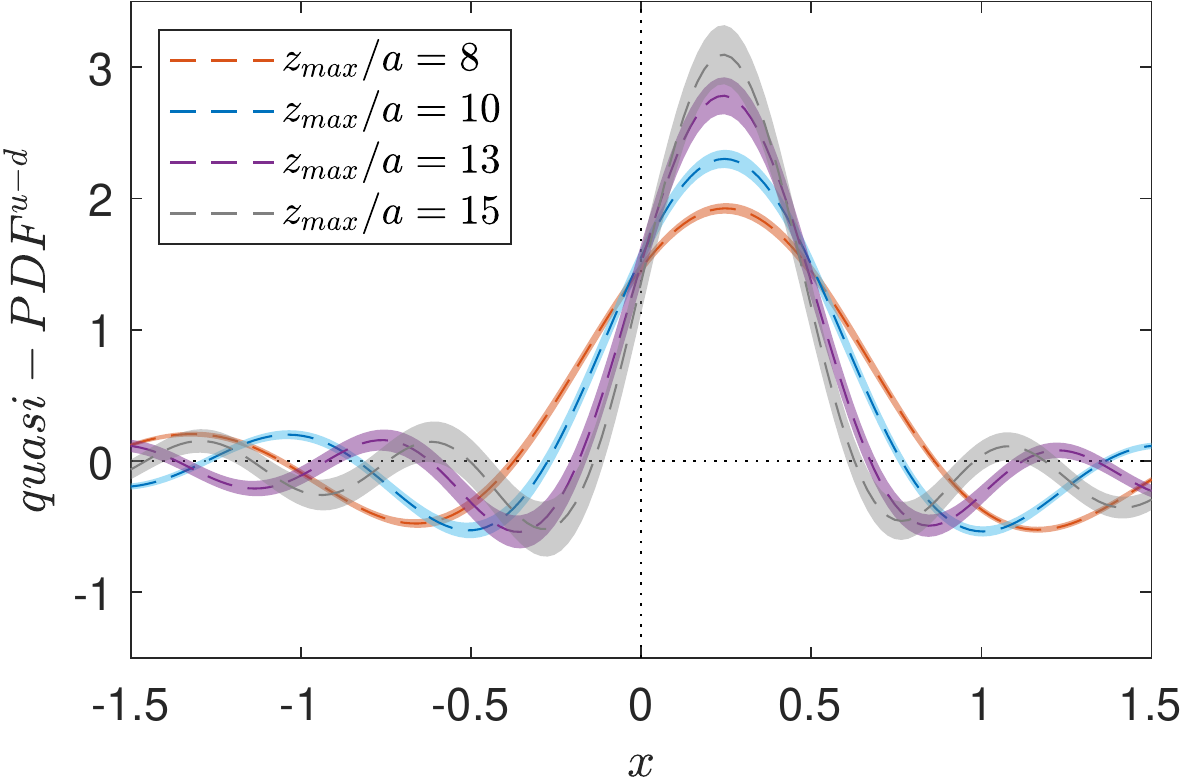}}\hspace{5mm}
 	\subfigure{\includegraphics[width=0.48\textwidth]{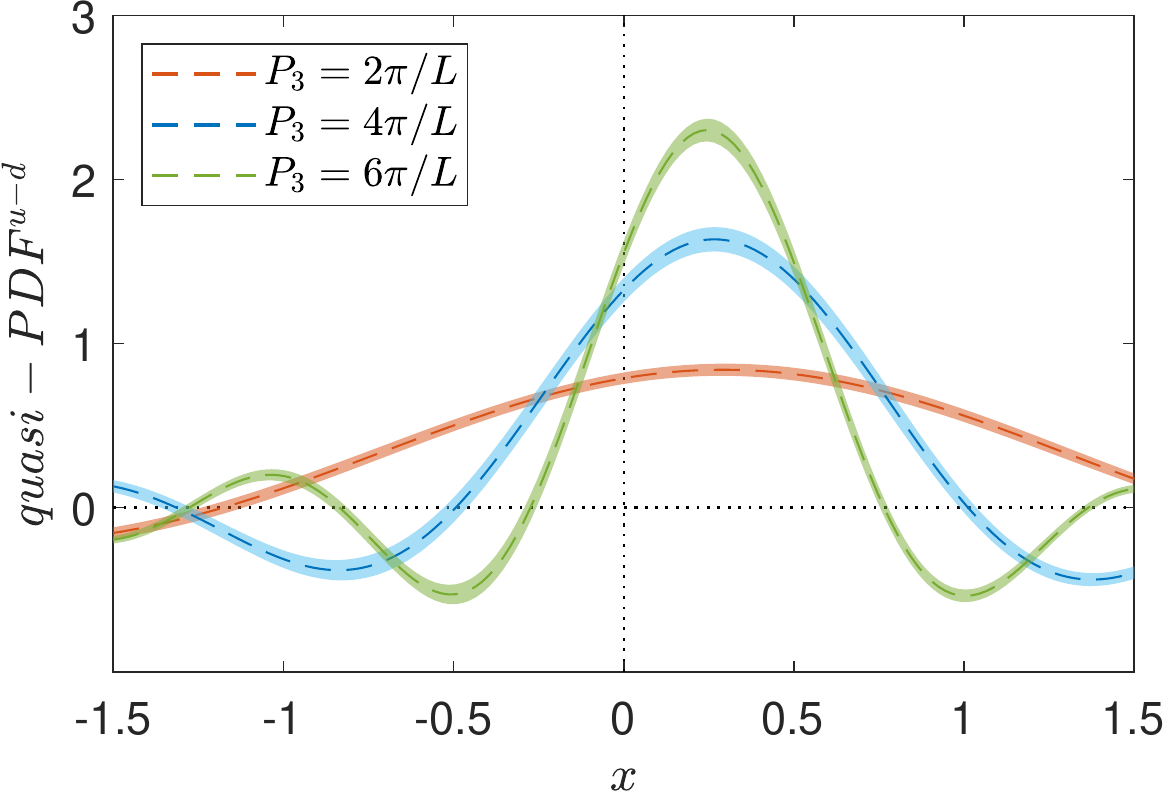}}
 	\caption{Quasi-PDFs corresponding to our renormalized matrix elements. Left panel: the dependence on the maximum length of the Wilson line taken in the Fourier transform, $z_{\rm max}$, for hadron boost $P_3=6\pi/L$. Right panel: the dependence on the hadron boost, for $P_3=2\pi/L,4\pi/L,6\pi/L$ and $z_{\rm max}/a=10$.}
 	\label{fig:quasi}
 \end{figure}

To obtain quasi-PDFs corresponding to our renormalized matrix elements $h^{\rm ren}(z,P_3)$, we perform a discrete Fourier transform,
\begin{equation}
\tilde{q}(x,P_3)=\frac{2P_3}{4\pi}\sum_{-z_{\rm max}}^{z_{\rm max}}e^{-ixP_3z} h^{\rm ren}(z,P_3),
\end{equation}
where $z_{\rm max}$ is the maximum length of the Wilson line used in the procedure.
We investigate the influence of the choice of $z_{\rm max}$ on the results in the left panel of Fig.~\ref{fig:quasi}.
We observe that all regions of $x$ are rather significantly affected by $z_{\rm max}$.
Thus, the choice of this cutoff value is delicate and indicates that further investigations are necessary.
Ideally, there should be no dependence on the value of $z_{\rm max}$.
This corresponds to a situation when both the real part and the imaginary part of the renormalized matrix elements decay to zero at some value of $|z|$ and remain zero for larger lengths of the Wilson line.
In practice, as can be seen in Fig.~\ref{renormalization matrix}, the real part for 10 stout smearing iterations becomes consistent with zero around $|z|/a=10$ and becomes slightly negative for larger values of $|z|$, with large uncertainties.
The imaginary part decays more slowly and it is non-zero for any $|z|\leq L/2$, for which we have computed the matrix elements.
Bare matrix elements are very close to zero at $|z|/a=L/2$.
However, at large Wilson line lengths, they are significantly enhanced by the large values of the renormalization functions.
Moreover, the renormalization functions at large $z$ are possibly subject to non-physical effects resulting from the perturbative conversion to the $\MMSb$ scheme.
In particular, we observe that the renormalization functions in this scheme have a non-vanishing imaginary part, relatively large especially at large $z$.
This is, clearly, a truncation effect resulting from the use of only one-loop perturbative formulae.
Also from the point of factorization of the quasi-PDF into the light-cone PDF, large values of $z$ should be avoided \cite{Ma:2017pxb,Ma:2014jla}.
With the current data, we opt for the choice of $z_{\rm max}/a=10$, where the real part of renormalized matrix elements becomes zero.
Nevertheless, in further studies, more advanced than our current exploratory study, it is desirable to reduce the uncertainty related to the choice of $z_{\rm max}$, by achieving a larger hadron boost and minimizing truncation effects in the perturbative conversion to the $\MMSb$ scheme.
Additionally, the discrete Fourier transform should be replaced by advanced reconstruction techniques, as advocated in Ref.~\cite{Karpie:2019eiq}.
These techniques provide a model-independent way of supplementing the lattice data, which are necessarily restricted to finite lengths of the Wilson line and to a finite, $\mathcal{O}(10-20)$, number of evaluations of the matrix elements.
 
In the right panel of Fig.~\ref{fig:quasi}, we show the quasi-PDFs from $z_{\rm max}/a=10$, for our three hadron boosts.
We observe rather large dependence on the particle momentum.
Obviously, the finite-momentum distributions are expected to depend visibly on the boost and conclusions about convergence to the infinite momentum frame can only be drawn after the matching procedure, which we discuss below.
After this procedure, we also expect that, provided the momentum is large enough, the support of the distributions becomes close to the canonical one, $x\in[-1,1]$.

The matching procedure relies on a factorization relation between the quasi-PDF and the light-cone PDF, Eq.~(\ref{eq:matching equation}), where the matching function $C(x/y,\mu/yP_3)$ is computed in perturbation theory.
This factorization holds up to power-suppressed corrections in $1/P_3^2$.
For the matching kernel, we use the one-loop expression introduced in Ref.~\cite{Alexandrou:2018pbm}, which is valid for a quasi-PDF expressed in the $\MMSb$ scheme and the action of which provides the light-cone distribution expressed in the $\MSb$ scheme.
For completeness, we write below the final formula and for a broader discussion, we refer to Ref.~\cite{Alexandrou:2019lfo}:
\begin{equation}
C\left(\xi, \frac{\xi \mu}{x P_{3}}\right)=\delta(1-\xi)+\frac{\alpha_{s}}{2 \pi} C_{F}\left\{\begin{array}{ll}{\left[\frac{1+\xi^{2}}{1-\xi} \ln \frac{\xi}{\xi-1}+1+\frac{3}{2 \xi}\right]_{+(1)}} & {\xi>1},
\\ {\left[\frac{1+\xi^{2}}{1-\xi} \ln \frac{x^{2} P_{3}^{2}}{\xi^{2} \mu^{2}}(4 \xi(1-\xi))-\frac{\xi(1+\xi)}{1-\xi}\right]_{+(1)}} & {0<\xi<1},
\\ {\left[-\frac{1+\xi^{2}}{1-\xi} \ln \frac{\xi}{\xi-1}-1+\frac{3}{2(1-\xi)}\right]_{+(1)}} & {\xi<0},\end{array}\right.
\end{equation}
where the plus prescription is defined as $\int d\xi [f(\xi)]_{+(1)}q(x/\xi,\mu)= \int d\xi f(\xi)\left[q(x/\xi,\mu)-q(x,\mu)\right]$.
We also remove $O(M^2/P_3^2)$ hadron mass corrections which were calculated in a closed form  in Ref.~\cite{Chen:2016utp}.

\begin{figure}[h]
 	\centering
 	\subfigure{\includegraphics[width=0.48\textwidth]{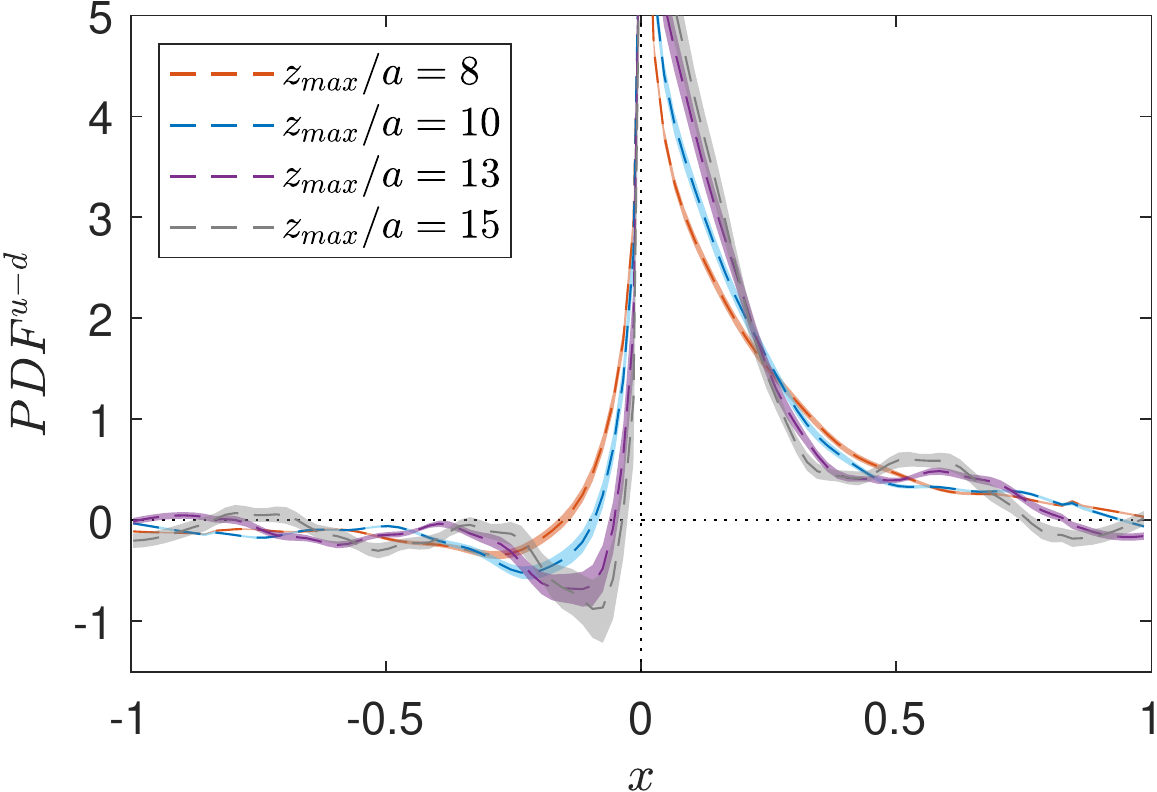}}\hspace{5mm}
 	\subfigure{\includegraphics[width=0.48\textwidth]{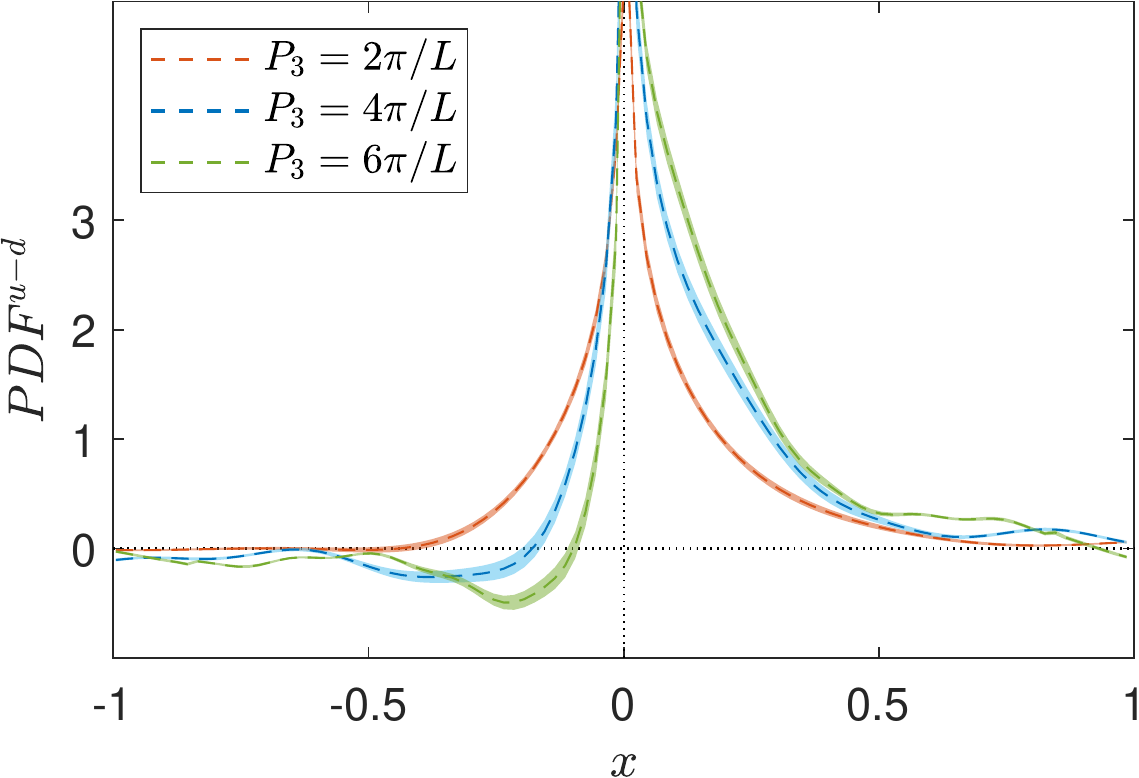}}
 	\caption{Matched PDFs obtained from the quasi-PDFs in Fig.~\ref{fig:quasi}. Left panel: the dependence on the maximum length of the Wilson line taken in the Fourier transform, $z_{\rm max}$, for hadron boost $P_3=6\pi/L$. Right panel: the dependence on the hadron boost, for $P_3=2\pi/L,4\pi/L,6\pi/L$ and $z_{\rm max}/a=10$.}
 	\label{fig:PDF}
\end{figure}

In Fig.~\ref{fig:PDF}, we present the light-cone PDFs, after matching and mass corrections. The width of the band corresponds to the statistical uncertainties.
In the left panel, we illustrate how the $z_{\rm max}$-dependence of quasi-PDFs propagates into the light-cone frame.
The matched PDFs with our choice of $z_{\rm max}/a=10$ are shown in the right panel, for the three different momenta.
For all momenta, the canonical support in the Bjorken-$x$ variable is approximately restored by the matching.
We observe a significant change from $P_3=2\pi/L$ to $P_3=4\pi/L$ and a considerably smaller one to $P_3=6\pi/L$.
The dependence on the hadron boost is largely reduced with respect to the boost dependence of quasi-PDFs (cf.\ the right panel of Fig.~\ref{fig:quasi}).
Nevertheless, at this level of statistical precision, the matching cannot account for the differences in momenta of quasi-PDFs.
Therefore, as also argued above, larger momenta are needed to make reliable contact to the light-cone distributions.
This requires significantly more computational resources than the ones used in the present study and thus, we will address it in the future.

\begin{figure}[h!]
\centerline{\includegraphics[width=0.48\textwidth]{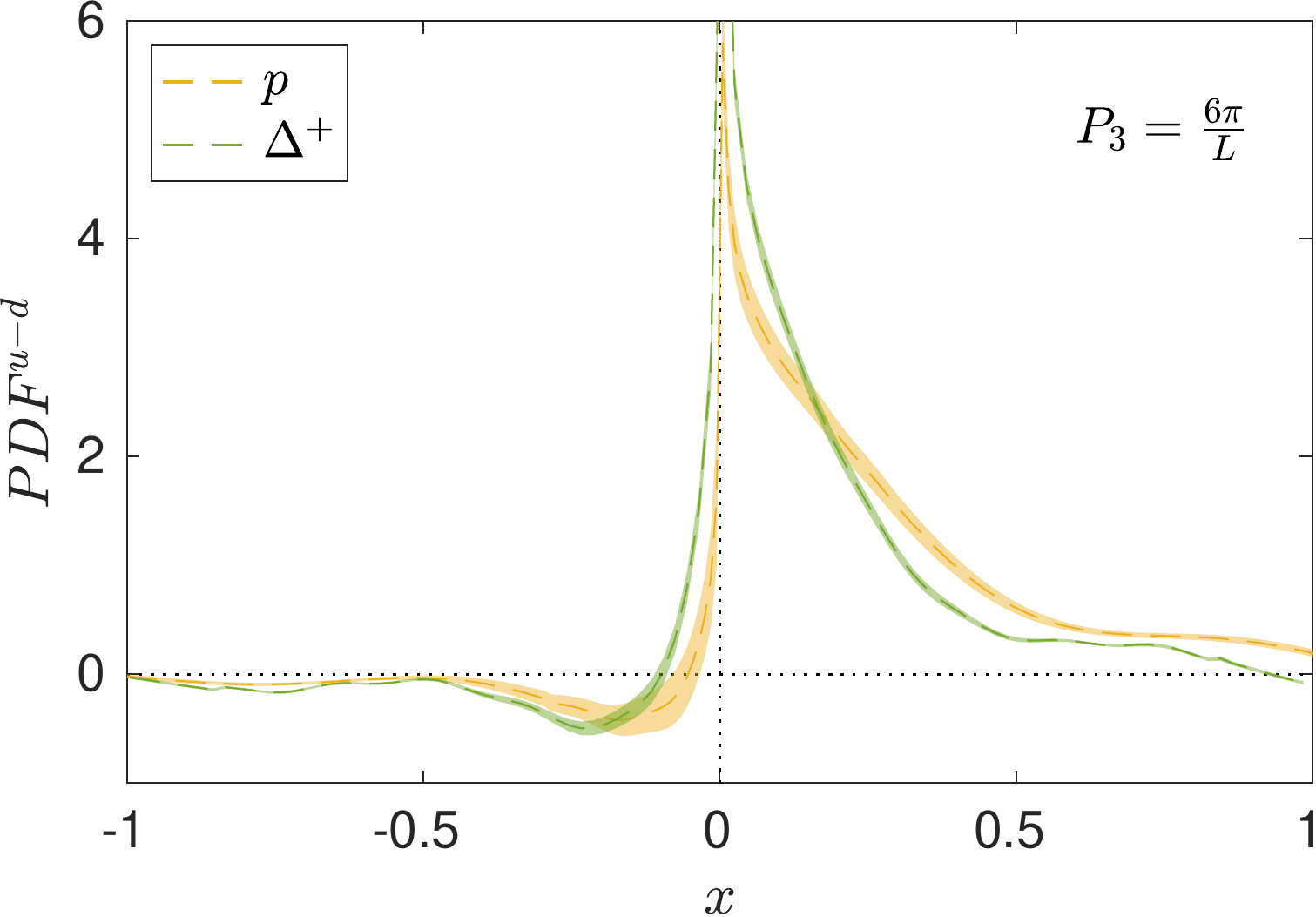}}
     \caption{Comparison between the unpolarized PDF for the proton (yellow) and $\Delta^+$ baryon (green), obtained for the ensemble at $m_\pi=270$~MeV and momentum boost $P_3=6\pi/L$. The same cutoff $z_{max}/a=10$ has been applied for both lattice results.}
     \label{fig:comparison_proton}
 \end{figure}
 
\begin{figure}
\centerline{\includegraphics[width=0.48\textwidth]{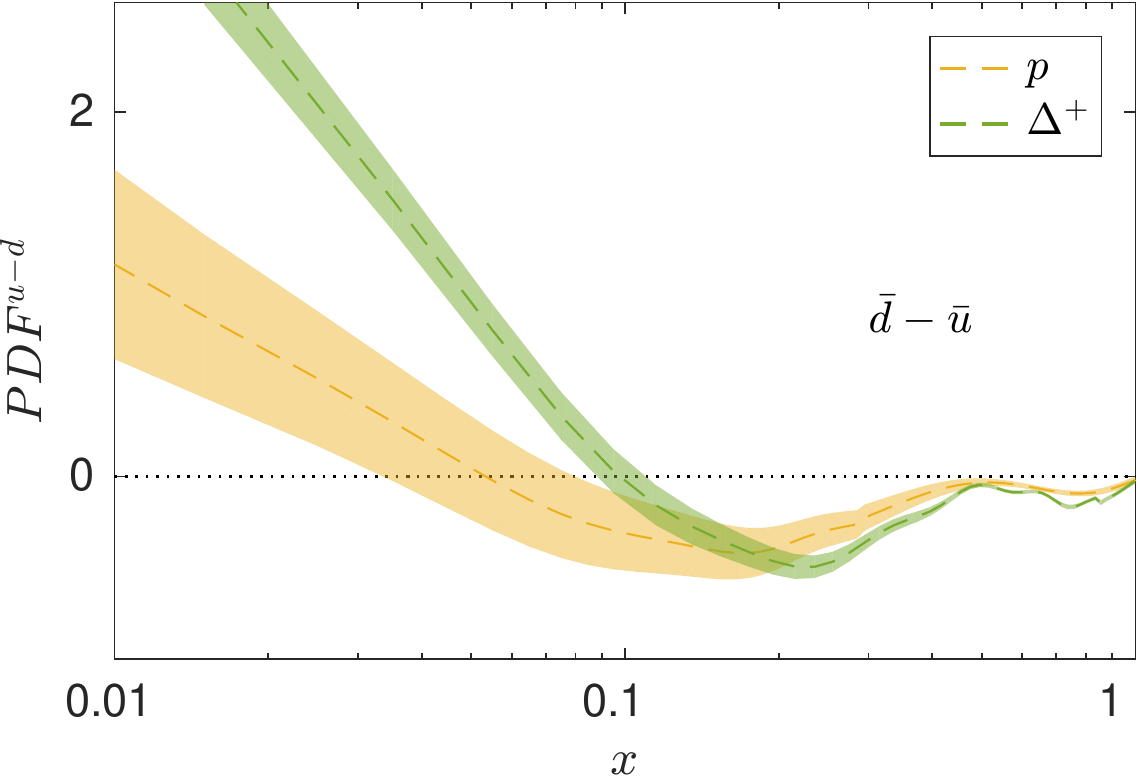}}
     \caption{Comparison between the $\overline{d}(x)-\overline{u}(x)$ asymmetry for the proton (yellow) and $\Delta^+$ baryon (green), obtained for the ensemble at $m_\pi=270$~MeV and momentum boost $P_3=6\pi/L$. The same cutoff $z_{max}/a=10$ has been applied for both lattice results. }
     \label{fig:comparison_proton_log}
 \end{figure}

In Fig.~\ref{fig:comparison_proton} we show our final result for the $\Delta$ baryon for our highest value of $P_3$, 
along with the corresponding results for the proton obtained on the same ensemble. Only statistical uncertainties are included in the bands. For the proton, we use 1688 measurements, which explains the larger errors as compared to the $\Delta$.
We recall that despite the fact that the pion mass is slightly smaller than the $\Delta - p$ mass difference, in our kinematical setup the $\Delta$ baryon is near threshold and, thus, assumed to be stable, as the decay is strongly suppressed. With this in mind, we make a first attempt to compare our results with the 
prediction of Ref.~\cite{Ethier:2018efr} for the $\overline{d}(x)-\overline{u}(x)$ asymmetry in the $\Delta+$ and in the proton. As can be seen in the negative $x$ region of Fig.~\ref{fig:comparison_proton} (antiquark region), $\overline{d}(x)-\overline{u}(x)$ is larger in the $\Delta^+$. For better visualization, we show in Fig.~\ref{fig:comparison_proton_log}
the asymmetry for both baryons using a logarithmic scale. It is interesting to observe that there is an indication of increased $\overline{d}(x)-\overline{u}(x)$ asymmetry for the $\Delta^+$ as compared to the proton. This is in line with what is expected according to Fig. 5 of Ref.~\cite{Ethier:2018efr}, despite the oscillations in the $x \gtrapprox 0.1$ region. However, further investigation of systematic uncertainties is required to arrive at reliable conclusions.

\section{Summary and outlook}
In this work, we performed a first lattice QCD calculation of the unpolarized isovector PDF of the $\Delta^+$ baryon using LaMET. Our main goal was to demonstrate the feasibility of extracting the PDFs of the $\Delta^+$ with high statistical accuracy, while the computational cost remains within reach.
We considered a non-physical situation of the $\Delta$ being 15-30\% heavier than in nature, which situates it very close to the $\Delta\rightarrow N\pi$ decay threshold. In such a case, the decay, even if allowed, is strongly suppressed and one can use the existing techniques of quasi-distributions to probe the $\Delta$ PDFs, in particular to access the $\overline{d}(x)-\overline{u}(x)$ asymmetry in this hadron. The latter is one of the main motivations for this study.
As calculated in Ref.~\cite{Ethier:2018efr}, the asymmetry should be larger in the $\Delta^+$ baryon than in the proton, becoming substantial as $x$ becomes smaller and as the $\Delta^+$ decay threshold is approached. Fig.~\ref{fig:comparison_proton_log}, the main outcome of our work, suggests such behavior, which is in the correct direction. To obtain reliable conclusions on the physical mechanism behind this asymmetry, further investigations of systematic uncertainties are required.

In this work, we computed the matrix elements of quasi-PDFs at three values of the hadron boost, 0.42 GeV, 0.83 GeV and 1.25 GeV. The matrix elements were renormalized non-perturbatively using the RI$^\prime$ scheme, and we used the modified matching coefficient, which ensures the particle number conservation in the matching procedure, to convert the quasi-PDF to the light-cone PDF in the $\overline{\rm MS}$ scheme at $\mu=2$ GeV. We also subtracted the $O(M^2/P_3^2)$ hadron mass corrections. 
Although the noise in the computations for the $\Delta$ baryon was expected to be larger than for the nucleon, we managed to show in this study that these difficulties can be overcome by standard techniques, using reasonable computational resources. This led to an extraction of the PDFs with an accuracy comparable to the one of the nucleon. 

Of course, our calculation needs to be improved in future work, in particular concerning possible volume and cut-off effects, which may be fundamental to address the problem with oscillations in the $x$ distributions which currently permeats our results. 
Another direction is the investigation of other systematic uncertainties, such as excited states effects. We plan to follow the methodology presented in Ref.~\cite{Alexandrou:2019lfo}, i.e.\ using three methods (plateau fits, two-state fits and the summation method) at the largest employed hadron boost.
A computation with higher values of $P_3$ and with a setup such that the $\Delta^+$ is even closer to the $\pi N$ decay channel is also one of our priorities, because of the possible observation of the rapid rise of  $\overline{d}(x)-\overline{u}(x)$ in the $\Delta^+$ when compared to the nucleon as the decay channel is approached.
We note that being close to the threshold is desirable from the point of view of testing the role of chiral symmetry according to the mechanisms of Ref.~\cite{Ethier:2018efr} and thus, this thread of work does not motivate working at the physical pion mass.
The latter could be interesting in its own right to provide statements about the physical $\Delta(1232)$ baryon, but, as we emphasized above, would require the development of the quasi-PDF formalism for unstable particles.

\noindent
\section{Acknowledgments} We would like to thank all members of the Extended Twisted Mass Collaboration 
for their constant and pleasant collaboration. Y.C., Y.L. and C.L. are supported in part by the Ministry of Science and Technology of China (MSTC) under 973 project "Systematic studies on light hadron spectroscopy", No. 2015CB856702 and the DFG and the NSFC through funds
 provided to the Sino-Germen CRC 110 ``Symmetries and the Emergence of Structure in QCD'', DFG grant no. TRR~110 and NSFC grant No. 11621131001. X.F. and S.X. are supported in part by NSFC of China under Grant No. 11775002. This work has also received funding from the European Union's Horizon 2020
research and innovation programme under the Marie Sk\l{}odowska-Curie grant agreement
No 642069 (HPC-LEAP). 
K.C.\ and A.S.\ were supported by National Science Centre (Poland) grant SONATA
BIS no.\ 2016/22/E/ST2/00013. F.S.\ was funded by DFG project number 392578569.
M.C. acknowledges financial support by the U.S. Department of Energy, Office of Nuclear Physics, within
the framework of the TMD Topical Collaboration. K.H. is financially supported by the Cyprus Research Promotion foundation under contract number POST-DOC/0718/0100.
The calculation was carried out on TianHe-3 (prototype) at Chinese National Supercomputer Center in Tianjin. It also used resources of
JUWELS at  the Juelich Supercomputing Centre,  under project id ECY00. The gauge configurations have been generated on the KNL (A2) Partition of
Marconi at CINECA, through the Prace project Pra13\_3304 "SIMPHYS".

\bibliography{reference}

\begin{thebibliography}{10}

\bibitem{Abdel-Rehim:2015owa}
A.~Abdel-Rehim {\em et~al.},
\newblock Phys. Rev. {\bf D92}, 114513 (2015), 1507.04936,
\newblock [Erratum: Phys. Rev.D93,no.3,039904(2016)].

\bibitem{Alexandrou:2019olr}
C.~Alexandrou {\em et~al.},
\newblock (2019), 1909.10744.

\bibitem{Alexandrou:2019brg}
C.~Alexandrou {\em et~al.},
\newblock (2019), 1909.00485.

\bibitem{Alexandrou:2019ali}
C.~Alexandrou {\em et~al.},
\newblock (2019), 1908.10706.

\bibitem{Alexandrou:2018sjm}
C.~Alexandrou {\em et~al.},
\newblock Phys. Rev. {\bf D100}, 014509 (2019), 1812.10311.

\bibitem{Oehm:2018jvm}
M.~Oehm {\em et~al.},
\newblock Phys. Rev. {\bf D99}, 014508 (2019), 1810.09743.

\bibitem{Ji:2013dva}
X.~Ji,
\newblock Phys. Rev. Lett. {\bf 110}, 262002 (2013), 1305.1539.

\bibitem{Ji:2014gla}
X.~Ji,
\newblock Sci. China Phys. Mech. Astron. {\bf 57}, 1407 (2014), 1404.6680.

\bibitem{Xiong:2013bka}
X.~Xiong, X.~Ji, J.-H. Zhang, and Y.~Zhao,
\newblock Phys.Rev. {\bf D90}, 014051 (2014), 1310.7471.

\bibitem{Ma:2014jla}
Y.-Q. Ma and J.-W. Qiu,
\newblock Phys. Rev. {\bf D98}, 074021 (2018), 1404.6860.

\bibitem{Briceno:2017cpo}
R.~A. Brice{\~n}o, M.~T. Hansen, and C.~J. Monahan,
\newblock Phys. Rev. {\bf D96}, 014502 (2017), 1703.06072.

\bibitem{Ma:2017pxb}
Y.-Q. Ma and J.-W. Qiu,
\newblock Phys. Rev. Lett. {\bf 120}, 022003 (2018), 1709.03018.

\bibitem{Chen:2016utp}
J.-W. Chen, S.~D. Cohen, X.~Ji, H.-W. Lin, and J.-H. Zhang,
\newblock Nucl. Phys. {\bf B911}, 246 (2016), 1603.06664.

\bibitem{Liu:1993cv}
K.-F. Liu and S.-J. Dong,
\newblock Phys. Rev. Lett. {\bf 72}, 1790 (1994), hep-ph/9306299.

\bibitem{Aglietti:1998ur}
U.~Aglietti {\em et~al.},
\newblock Phys. Lett. {\bf B441}, 371 (1998), hep-ph/9806277.

\bibitem{Detmold:2005gg}
W.~Detmold and C.~J.~D. Lin,
\newblock Phys. Rev. {\bf D73}, 014501 (2006), hep-lat/0507007.

\bibitem{Braun:2007wv}
V.~Braun and D.~Mueller,
\newblock Eur. Phys. J. {\bf C55}, 349 (2008), 0709.1348.

\bibitem{Radyushkin:2016hsy}
A.~Radyushkin,
\newblock Phys. Lett. {\bf B767}, 314 (2017), 1612.05170.

\bibitem{Radyushkin:2017cyf}
A.~V. Radyushkin,
\newblock Phys. Rev. {\bf D96}, 034025 (2017), 1705.01488.

\bibitem{Orginos:2017kos}
K.~Orginos, A.~Radyushkin, J.~Karpie, and S.~Zafeiropoulos,
\newblock Phys. Rev. {\bf D96}, 094503 (2017), 1706.05373.

\bibitem{Radyushkin:2017lvu}
A.~V. Radyushkin,
\newblock Phys. Lett. {\bf B781}, 433 (2018), 1710.08813.

\bibitem{Chambers:2017dov}
A.~J. Chambers {\em et~al.},
\newblock Phys. Rev. Lett. {\bf 118}, 242001 (2017), 1703.01153.

\bibitem{Karpie:2018zaz}
J.~Karpie, K.~Orginos, and S.~Zafeiropoulos,
\newblock JHEP {\bf 11}, 178 (2018), 1807.10933.

\bibitem{Radyushkin:2018cvn}
A.~Radyushkin,
\newblock Phys. Rev. {\bf D98}, 014019 (2018), 1801.02427.

\bibitem{Bali:2018spj}
G.~S. Bali {\em et~al.},
\newblock Phys. Rev. {\bf D98}, 094507 (2018), 1807.06671.

\bibitem{Detmold:2018kwu}
W.~Detmold, I.~Kanamori, C.~J.~D. Lin, S.~Mondal, and Y.~Zhao,
\newblock {Moments of pion distribution amplitude using operator product
  expansion on the lattice},
\newblock 2018, 1810.12194.

\bibitem{Sufian:2019bol}
R.~S. Sufian {\em et~al.},
\newblock Phys. Rev. {\bf D99}, 074507 (2019), 1901.03921.

\bibitem{Liang:2019frk}
XQCD, J.~Liang, T.~Draper, K.-F. Liu, A.~Rothkopf, and Y.-B. Yang,
\newblock (2019), 1906.05312.

\bibitem{Joo:2019jct}
B.~Joó {\em et~al.},
\newblock JHEP {\bf 12}, 081 (2019), 1908.09771.

\bibitem{Joo:2019bzr}
B.~Joó {\em et~al.},
\newblock Phys. Rev. {\bf D100}, 114512 (2019), 1909.08517.

\bibitem{Lin:2014zya}
H.-W. Lin, J.-W. Chen, S.~D. Cohen, and X.~Ji,
\newblock Phys. Rev. {\bf D91}, 054510 (2015), 1402.1462.

\bibitem{Alexandrou:2015rja}
C.~Alexandrou {\em et~al.},
\newblock Phys. Rev. {\bf D92}, 014502 (2015), 1504.07455.

\bibitem{Alexandrou:2016jqi}
C.~Alexandrou {\em et~al.},
\newblock Phys. Rev. {\bf D96}, 014513 (2017), 1610.03689.

\bibitem{Zhang:2017bzy}
J.-H. Zhang, J.-W. Chen, X.~Ji, L.~Jin, and H.-W. Lin,
\newblock Phys. Rev. {\bf D95}, 094514 (2017), 1702.00008.

\bibitem{Lin:2017ani}
LP3, H.-W. Lin, J.-W. Chen, T.~Ishikawa, and J.-H. Zhang,
\newblock Phys. Rev. {\bf D98}, 054504 (2018), 1708.05301.

\bibitem{Chen:2017gck}
LP3, J.-H. Zhang {\em et~al.},
\newblock Nucl. Phys. {\bf B939}, 429 (2019), 1712.10025.

\bibitem{Chen:2018fwa}
J.-H. Zhang {\em et~al.},
\newblock Phys. Rev. {\bf D100}, 034505 (2019), 1804.01483.

\bibitem{Alexandrou:2018pbm}
C.~Alexandrou {\em et~al.},
\newblock Phys. Rev. Lett. {\bf 121}, 112001 (2018), 1803.02685.

\bibitem{Lin:2018qky}
H.-W. Lin {\em et~al.},
\newblock Phys. Rev. Lett. {\bf 121}, 242003 (2018), 1807.07431.

\bibitem{Liu:2018uuj}
Y.-S. Liu {\em et~al.},
\newblock (2018), 1807.06566.

\bibitem{Fan:2018dxu}
Z.-Y. Fan, Y.-B. Yang, A.~Anthony, H.-W. Lin, and K.-F. Liu,
\newblock Phys. Rev. Lett. {\bf 121}, 242001 (2018), 1808.02077.

\bibitem{Alexandrou:2018eet}
C.~Alexandrou {\em et~al.},
\newblock Phys. Rev. {\bf D98}, 091503 (2018), 1807.00232.

\bibitem{Izubuchi:2019lyk}
T.~Izubuchi {\em et~al.},
\newblock Phys. Rev. {\bf D100}, 034516 (2019), 1905.06349.

\bibitem{Ishikawa:2017faj}
T.~Ishikawa, Y.-Q. Ma, J.-W. Qiu, and S.~Yoshida,
\newblock Phys. Rev. {\bf D96}, 094019 (2017), 1707.03107.

\bibitem{Ji:2017oey}
X.~Ji, J.-H. Zhang, and Y.~Zhao,
\newblock Phys. Rev. Lett. {\bf 120}, 112001 (2018), 1706.08962.

\bibitem{Zhang:2018diq}
J.-H. Zhang, X.~Ji, A.~Schäfer, W.~Wang, and S.~Zhao,
\newblock Phys. Rev. Lett. {\bf 122}, 142001 (2019), 1808.10824.

\bibitem{Li:2018tpe}
Z.-Y. Li, Y.-Q. Ma, and J.-W. Qiu,
\newblock Phys. Rev. Lett. {\bf 122}, 062002 (2019), 1809.01836.

\bibitem{Alexandrou:2017huk}
C.~Alexandrou {\em et~al.},
\newblock Nucl. Phys. {\bf B923}, 394 (2017), 1706.00265.

\bibitem{Green:2017xeu}
J.~Green, K.~Jansen, and F.~Steffens,
\newblock Phys. Rev. Lett. {\bf 121}, 022004 (2018), 1707.07152.

\bibitem{Ji:2015qla}
X.~Ji, A.~Sch{\"a}fer, X.~Xiong, and J.-H. Zhang,
\newblock Phys. Rev. {\bf D92}, 014039 (2015), 1506.00248.

\bibitem{Xiong:2015nua}
X.~Xiong and J.-H. Zhang,
\newblock Phys. Rev. {\bf D92}, 054037 (2015), 1509.08016.

\bibitem{Wang:2017qyg}
W.~Wang, S.~Zhao, and R.~Zhu,
\newblock Eur. Phys. J. {\bf C78}, 147 (2018), 1708.02458.

\bibitem{Stewart:2017tvs}
I.~W. Stewart and Y.~Zhao,
\newblock Phys. Rev. {\bf D97}, 054512 (2018), 1709.04933.

\bibitem{Izubuchi:2018srq}
T.~Izubuchi, X.~Ji, L.~Jin, I.~W. Stewart, and Y.~Zhao,
\newblock Phys. Rev. {\bf D98}, 056004 (2018), 1801.03917.

\bibitem{Liu:2018hxv}
Y.-S. Liu {\em et~al.},
\newblock (2018), 1810.05043.

\bibitem{Cichy:2018mum}
K.~Cichy and M.~Constantinou,
\newblock Adv. High Energy Phys. {\bf 2019}, 3036904 (2019), 1811.07248.

\bibitem{Alexandrou:2019lfo}
C.~Alexandrou {\em et~al.},
\newblock Phys. Rev. {\bf D99}, 114504 (2019), 1902.00587.

\bibitem{Ethier:2018efr}
J.~J. Ethier, W.~Melnitchouk, F.~Steffens, and A.~W. Thomas,
\newblock (2018), 1809.06885.

\bibitem{Thomas:2000ny}
A.~W. Thomas, W.~Melnitchouk, and F.~M. Steffens,
\newblock Phys. Rev. Lett. {\bf 85}, 2892 (2000), hep-ph/0005043.

\bibitem{Chen:2001eg}
J.-W. Chen and X.-d. Ji,
\newblock Phys. Lett. {\bf B523}, 107 (2001), hep-ph/0105197.

\bibitem{Collins:2011zzd}
J.~Collins,
\newblock Camb. Monogr. Part. Phys. Nucl. Phys. Cosmol. {\bf 32}, 1 (2011).

\bibitem{Constantinou:2017sej}
M.~Constantinou and H.~Panagopoulos,
\newblock Phys. Rev. {\bf D96}, 054506 (2017), 1705.11193.

\bibitem{Alexandrou:2018egz}
C.~Alexandrou {\em et~al.},
\newblock Phys. Rev. {\bf D98}, 054518 (2018), 1807.00495.

\bibitem{Frezzotti:2000nk}
Alpha, R.~Frezzotti, P.~A. Grassi, S.~Sint, and P.~Weisz,
\newblock JHEP {\bf 08}, 058 (2001), hep-lat/0101001.

\bibitem{Frezzotti:2003ni}
R.~Frezzotti and G.~C. Rossi,
\newblock JHEP {\bf 08}, 007 (2004), hep-lat/0306014.

\bibitem{Sheikholeslami:1985ij}
B.~Sheikholeslami and R.~Wohlert,
\newblock Nucl. Phys. {\bf B259}, 572 (1985).

\bibitem{Iwasaki:1985we}
Y.~Iwasaki,
\newblock Nucl. Phys. {\bf B258}, 141 (1985).

\bibitem{Alexandrou:2008tn}
European Twisted Mass, C.~Alexandrou {\em et~al.},
\newblock Phys. Rev. {\bf D78}, 014509 (2008), 0803.3190.

\bibitem{Bali:2016lva}
G.~S. Bali, B.~Lang, B.~U. Musch, and A.~Schäfer,
\newblock Phys. Rev. {\bf D93}, 094515 (2016), 1602.05525.

\bibitem{Gusken:1989qx}
S.~Gusken,
\newblock Nucl. Phys. Proc. Suppl. {\bf 17}, 361 (1990).

\bibitem{Alexandrou:1992ti}
C.~Alexandrou, S.~Gusken, F.~Jegerlehner, K.~Schilling, and R.~Sommer,
\newblock Nucl. Phys. {\bf B414}, 815 (1994), hep-lat/9211042.

\bibitem{Albanese:1987ds}
APE, M.~Albanese {\em et~al.},
\newblock Phys. Lett. {\bf B192}, 163 (1987).

\bibitem{Dotsenko:1979wR}
V.~S. Dotsenko and S.~N. Vergeles,
\newblock Nucl. Phys. {\bf B169}, 527 (1980).

\bibitem{Brandt:1981kf}
R.~A. Brandt, F.~Neri, and M.-a. Sato,
\newblock Phys. Rev. {\bf D24}, 879 (1981).

\bibitem{Martinelli:1994ty}
G.~Martinelli, C.~Pittori, C.~T. Sachrajda, M.~Testa, and A.~Vladikas,
\newblock Nucl. Phys. {\bf B445}, 81 (1995), hep-lat/9411010.

\bibitem{Constantinou:2010gr}
ETM, M.~Constantinou {\em et~al.},
\newblock JHEP {\bf 08}, 068 (2010), 1004.1115.

\bibitem{Alexandrou:2015sea}
ETM, C.~Alexandrou, M.~Constantinou, and H.~Panagopoulos,
\newblock Phys. Rev. {\bf D95}, 034505 (2017), 1509.00213.

\bibitem{Gockeler:1998ye}
M.~Gockeler {\em et~al.},
\newblock Nucl. Phys. {\bf B544}, 699 (1999), hep-lat/9807044.

\bibitem{Constantinou:2014fka}
M.~Constantinou {\em et~al.},
\newblock Phys. Rev. {\bf D91}, 014502 (2015), 1408.6047.

\bibitem{Morningstar:2003gk}
C.~Morningstar and M.~J. Peardon,
\newblock Phys. Rev. {\bf D69}, 054501 (2004), hep-lat/0311018.

\bibitem{Karpie:2019eiq}
J.~Karpie, K.~Orginos, A.~Rothkopf, and S.~Zafeiropoulos,
\newblock JHEP {\bf 04}, 057 (2019), 1901.05408.

\end{thebibliography}

\end{document}